\documentclass[aps,prb,twocolumn,superscriptaddress,floatfix,showpacs,10pt,letterpaper]{revtex4-1}

\pdfoutput=1
\usepackage{graphicx, bm, amsmath, amsfonts,amssymb,amsthm}
\usepackage{subfigure, color}
\usepackage{ulem}
\newcommand{\be}{\begin{equation}}
\newcommand{\ee}{\end{equation}}
\newcommand{\Om}{\Omega}
\newcommand{\om}{\omega}
\newcommand{\al}{\alpha}
\newcommand{\bt}{\beta}
\newcommand{\cH}{H}
\newcommand{\cC}{C}
\newcommand{\cB}{B}
\newcommand{\cZ}{Z}
\newcommand{\zt}{\mathbb{Z}_2}
\newcommand{\ie}{i.e.}
\newcommand{\ba}{\begin{eqnarray} }
\newcommand{\ea}{\end{eqnarray} }
\newcommand{\n}{\nonumber \\ }
\newcommand{\mac}{\mathcal}

\newcommand{\ket}[1]{|#1 \rangle}
\newcommand{\bra}[1]{\langle #1 |}

\newcommand{\mz}{{\mathbb{Z}}}
\newcommand{\calH}{{\cal{H}}}

\DeclareTextFontCommand{\emph}{\it}
\begin{document}

\graphicspath{{figs/}}

\begin{titlepage}

\title{Anomalous Symmetry Fractionalization and Surface Topological Order}

\author{Xie Chen}
\affiliation{Department of Physics and Institute for Quantum Information and Matter, California Institute of Technology, Pasadena, CA 91125, USA}
\affiliation{Department of Physics, University of California, Berkeley, CA, 94720, USA}
\author{F. J. Burnell}
\affiliation{Department of Physics and Astronomy, University of Minnesota, Minneapolis, MN 55455, USA}
\author{Ashvin Vishwanath}
\affiliation{Department of Physics, University of California, Berkeley, CA 94720, USA.}
\affiliation{Materials Science Division, Lawrence Berkeley National Laboratories, Berkeley, CA 94720, USA.}
\author{Lukasz Fidkowski}
\affiliation{Department of Physics and Astronomy, Stony Brook University, Stony Brook, NY 11794-3800, USA.}

\begin{abstract}
In addition to possessing fractional statistics, anyon excitations of a 2D topologically ordered state can realize symmetry in distinct ways
, leading to a variety of symmetry enriched topological (SET) phases. While the symmetry fractionalization must be consistent with the fusion and braiding rules of the anyons, not all ostensibly consistent symmetry fractionalizations can be realized in 2D systems.  Instead, certain `anomalous' SETs can only occur on the surface of a 3D symmetry protected topological (SPT) phase. In this paper we describe a procedure for determining whether an SET of a discrete,  onsite, unitary symmetry group $G$ is anomalous or not.  The basic idea is to gauge the symmetry and expose the anomaly as an obstruction to a consistent topological theory combining both the original anyons and the gauge fluxes. Utilizing a result of Etingof, Nikshych, and Ostrik, we point out that a class of obstructions are captured by the fourth cohomology group $H^4( G, \,U(1))$, which also precisely labels the set of 3D SPT phases, with symmetry group $G$. We thus establish a general bulk-boundary correspondence between the anomalous SET and the 3d bulk SPT whose surface termination realizes it.  We illustrate this idea using the chiral spin liquid ($U(1)_2$) topological order with a reduced symmetry $\mz_2 \times \mz_2 \subset SO(3)$, which can act on the semion quasiparticle in an anomalous way.  We construct exactly solved  3d SPT models realizing the anomalous surface terminations, and demonstrate that they are non-trivial by computing three loop braiding statistics.  Possible extensions to anti-unitary symmetries are also discussed.

\end{abstract}


\maketitle

\end{titlepage}

\section{Introduction}

Recently it has been realized that gapped phases can be distinguished
on the basis of symmetry even when that symmetry is unbroken. Short
range entangled phases of this form - dubbed `symmetry protected
topological' (SPTs) - have been classified using group cohomology\cite{Chen2012,Chen2013}.
However, in two and higher dimensions there are also long range entangled phases
supporting fractionalized excitations (anyons), and it is an interesting problem to classify these phases
in the presence of a symmetry\cite{Wen2002,Levin2012,Essin2013,Mesaros2013,Hung2013,Lu2013,Fidkowski2014, Maissam2014}. In two dimensions, one approach is to distinguish such `symmetry enriched topological' (SET) phases on the basis of symmetry fractionalization on the anyons (note that this approach applies only when the symmetry does not permute the anyons)\cite{Wen2002,Essin2013}.  For example, a ${\mathbb Z}_2$ spin liquid with the gauge charge excitations carrying spin $1/2$ represents a different SET phase from the one where the gauge charge carries no spin. In two dimensions, all possible ways of assigning fractional symmetry quantum numbers to anyons which are compatible with their fusion and braiding rules can be enumerated \cite{Essin2013,Fidkowski2014}, and one may use this as a basis for classifying SETs.

However, it is not clear that just because a certain assignment of
fractional symmetry representations is compatible with all fusion
and braiding rules, it must necessarily be realizable by a two dimensional Hamiltonian. 
Previous works have put forward several putative SETs whose assignments are in fact 
\emph{anomalous}, and incompatible with any 2D symmetric physical
realization\cite{Vishwanath2013,Wang2013,Burnell2013,Fidkowski2013,Metlitski2013,Bonderson2013,Wang2013a,Chen2013a}. In these examples, time reversal symmetry is involved and the anomaly is usually exposed by showing that the SET must be chiral when realized in 2D, which necessarily breaks time reversal symmetry. In this paper, we focus on SETs with unitary discrete symmetries and discuss a general way to detect anomalies in them.  We start with the simplest example of this type, based on the topological order of a $\nu=1/2$ bosonic fractional quantum Hall effect (or chiral spin liquid), namely $U(1)_2$, and symmetry $G=\mz_2 \times \mz_2$.  While these choices certainly allow for a non-anomalous 2d SET, namely the Kalmeyer-Laughlin chiral spin liquid, with $\mz_2 \times \mz_2$ thought of as the subgroup of $180$ degree spatial rotations around the principal axes in $SO(3)$, we will show that they also allow three anomalous SETs, which we dub `anomalous projective semion' theories.

The  inconsistency in the anomalous projective semion theory is exposed when we try to gauge the $\mathbb{Z}_{2}\times\mathbb{Z}_{2}$ symmetry. If such an SET can be realized in a purely 2D symmetric model, this gauging process should result in a consistent larger topological theory that includes, in addition to the semion, gauge charges and gauge fluxes of $\mathbb{Z}_{2}\times\mathbb{Z}_{2}$.  However, as we will show, such an extension fails for the anomalous projective semion theories because braiding and fusion rules cannot be consistently defined for the gauge fluxes and the semion.  As we describe in more detail below, this failure can be seen as an irreparable violation of the pentagon equations for the gauge fluxes.  Thus the anomalous projective semion theories are impossible in 2D.  On the other hand, we show, by constructing an exactly solved model, that they can be realized at the surface of a 3D SPT with $\mathbb{Z}_{2}\times\mathbb{Z}_{2}$ symmetry.

Our exactly solved 3D lattice model is based on a general prescription due to Walker and Wang\cite{Walker12}, which essentially bootstraps a given 2d topological order into a 3D bulk Hamiltonian that realizes this topological order at its surface.  The original ``Walker-Wang" models of Refs. \onlinecite{Walker12, vonkeyserlingk13} have a trivial 3D bulk, which is a prerequisite for a 3D SPT.  To construct an actual SPT from these models, we must include additional degrees of freedom that transform as certain (linear) unitary representations of $\mz_2 \times \mz_2$, in such a way that the surface semion excitation transforms in the desired projective representation of the symmetry.  This `decorated' Walker-Wang model is necessarily a non-trivial SPT, since it realizes a surface that cannot exist on its own in 2D.  However, for completeness, we also compute the three loop braiding statistics introduced in \onlinecite{Wang2014a, Jiang2014} as a diagnostic of 3D SPT phases, and find it to be non-trivial.

The anomalous projective semion theories represent only the simplest examples of anomalous SETs, but the method discussed in this paper is generally applicable to any topological order and discrete unitary onsite global symmetry.  The mathematics underlying this method of anomaly detection, developed by Etingof et al.\cite{Etingof2010}, studies the problem of $G$-extensions of fusion categories.  Such $G$-extensions are constructed in stages by specifying certain data relating to how the anyons transform under the symmetry, and at each stage there is a potential obstruction to being able to continue the process.  It is gratifying that the same $H^4(G,U(1))$ used to classify 3d SPTs arises in a purely algebraic way as just such an obstruction - specifically, an obstruction to the pentagon equation for flux fusion rules.  We give some details about how to connect this algebraic approach to the physics (see also Refs. \onlinecite{Fidkowski2014, Maissam2014}).

The paper is structured as follows: in section \ref{anomaly}, we introduce the anomalous projective semion theories and show that gauging the $\mathbb{Z}_{2}\times\mathbb{Z}_{2}$ symmetry leads to inconsistencies; in section \ref{WW} we present solvable 3D lattice models that realize the 3D bulk SPT with an anomalous projective semion surface state; in section \ref{3loopbraiding} we make a connection to other 3d SPT approaches, in particular computing the three loop braiding in our exactly solved model and giving a non-linear sigma model 3d SPT construction of our anomalous surfaces; in section \ref{O5} we give a non-linear sigma model construction of our SPT and surface; in section \ref{disc} we summarize our findings and discuss future directions, including the incorporation of time reversal symmetry into this formalism.

\section{Anomaly of the projective semion model in 2D}
\label{anomaly}

\subsection{The `projective semion' model}
\label{model}

The `anomalous projective semion' model we consider is a variant of the Kalmeyer-Laughlin chiral spin liquid (CSL) \cite{Kalmeyer1987}. The Kalmeyer-Laughlin CSL can of course be realized in 2D with the explicit construction of Ref. \onlinecite{Kalmeyer1987}. However, by slightly modifying the way spin rotation symmetry acts on the semion, we obtain an anomalous theory.

First, recall the setup for the Kalmeyer Laughlin CSL. The degrees of freedom are spin-$1/2$'s on a lattice. We will not be concerned with the precise form of the Hamiltonian, but only note that it is spin rotation ($SO(3)$) invariant. Thus we can think of the CSL as an $SO(3)$ symmetry enriched topological (SET) phase. The chiral topological order is the same as the $\nu=1/2$ bosonic fractional quantum Hall state, which can be described by the $K=2$ Chern-Simons gauge theory
\be
\mathcal{L} = 2 \frac{\epsilon_{\mu\nu\lambda}}{4\pi} \ a_{\mu}\partial_{\nu}a_{\lambda}
\ee
There is one non-trivial anyon, a semion $s$ which induces a phase factor of $-1$ when going around another semion. Two semions fuse into the a trivial quasi-particle, which we denote as $I$.
Moreover each semion carries a spin-$1/2$ under the $SO(3)$ symmetry. The CSL is then a nontrivial SET because $s$ carries a projective representation of $SO(3)$.\cite{Note1}
The precise definition of projective representation is given in appendix \ref{Gcoh}. Such an SET theory is of course realizable in 2D and thus not anomalous.

In order to get an anomalous theory, first reduce the symmetry to the discrete subgroup of $180$ degree rotations about the $x,y,$ and $z$ axes, which form a $\mathbb{Z}_{2}\times\mathbb{Z}_{2}$ subgroup of $SO(3)$. We denote the group elements as $g_x$, $g_y$ and $g_z$.
The CSL is of course also an SET of this reduced symmetry group. Each semion carries half charge for all three $\mathbb{Z}_2$ transformations, because $360$ degree rotation of a spin-$1/2$ always results in a phase factor of $-1$. Moreover, the three $\mathbb{Z}_2$ transformations anti-commute with each other and can be represented as
\be
\text{CSL:} \ \  g_x = i\sigma_x, g_y = i\sigma_y, g_z = i\sigma_z
\label{CSL}
\ee

However, now there are other possible SETs, because the semion can now carry {\em{arbitrary}} half integral charges of the $\mathbb{Z}_2$ symmetries. For example, the semion can carry integral charge under $g_x$ and $g_y$ but half integral charge under $g_z$. Indeed, we can have three variants of the CSL, which we call the `anomalous projective semion' models, where the $\mathbb{Z}_{2}\times\mathbb{Z}_{2}$ on the semions can be represented as
\be
\begin{array}{l}
\text{Anom. proj. semion X:} \ \  g_x = i\sigma_x, g_y = \sigma_y, g_z = \sigma_z \\
\text{Anom. proj. semion Y:} \ \  g_x = \sigma_x, g_y = i\sigma_y, g_z = \sigma_z \\
\text{Anom. proj. semion Z:} \ \  g_x = \sigma_x, g_y = \sigma_y, g_z = i\sigma_z
\end{array}
\label{ps}
\ee

If we take $\sigma_x$ and $\sigma_y$ as the generators of $\mathbb{Z}_2 \times \mathbb{Z}_2$, then theories X, Y, and Z are simply ones where the semion carries a half charge under either the first, second, or both of these generators, respectively.   The addition of such half charges to the CSL seems completely harmless.  Indeed, note that it is compatible with the fusion rule of the semion: two semions fuse to a trivial quasiparticle, and two identical copies of any combination of half charges and spin 1/2's always fuse to an integral (non-fractionalized) representation of $\mathbb{Z}_2 \times \mathbb{Z}_2$. In fact, topological theories with the semion carrying half charges (in $\nu=1/2$ fractional quantum Hall states) and spin-$1/2$'s (in CSL) have both been identified in explicit models in 2D.  However, as we are going to show in the next section, the anomalous projective semion theories defined through equation \ref{ps} are not realizable in purely 2D systems with $\mathbb{Z}_{2}\times\mathbb{Z}_{2}$ symmetry.

\subsection{Projective fusion rules of symmetry defects}
\label{proj_fusion}

The anomaly in the theories defined by equation \ref{ps} will be exposed when we try to gauge the $\mathbb{Z}_2 \times \mathbb{Z}_2$ symmetry.  Since this is a discrete symmetry, this amounts to introducing gauge fluxes $\Om_x$, $\Om_y$ and $\Om_z$ and the corresponding gauge charges.  Here we will examine the first step in such a gauging process, in which symmetry defects -- i.e. confined versions of the gauge fluxes --  are introduced.   (We will nonetheless refer to these defects as fluxes).  The non-trivial projective action of $\mathbb{Z}_2 \times \mathbb{Z}_2$ on the semion over-constrains the fusion rules of the fluxes, leading to an inconsistency, as we show in this section.  Specifically, we show that {\emph{the gauge fluxes have a `projective'  fusion rule up to a semion.}} 

First, note that we can always bind a semion to any flux $\Om$, so each $\Om$ actually contains two topological superselection sectors.  One might try to label one of the sectors as the `vacuum' flux sector $\Om$ and the other as the `semion' flux sector $s\Om$, although there is no canonical way to choose which one should be labeled as the vacuum; it is only the difference between the two sectors that matters.  Interesting things happen when we consider the fusion rules of the fluxes.  Normally, one would expect for example $\Om_i \times \Om_i = I$ ($I$ denotes the vacuum) and $\Om_x \times \Om_y = \Om_z$ due to the structure of the symmetry group. However, due to the existence of two sectors we might actually get $\Om_i \times \Om_i = s$ and $\Om_x \times \Om_y = s\Om_z$. That is, the gauge fluxes must fuse in the expected way only up to an additional semion.

\begin{figure}[htbp]
\begin{center}
\includegraphics[width=7.0cm]{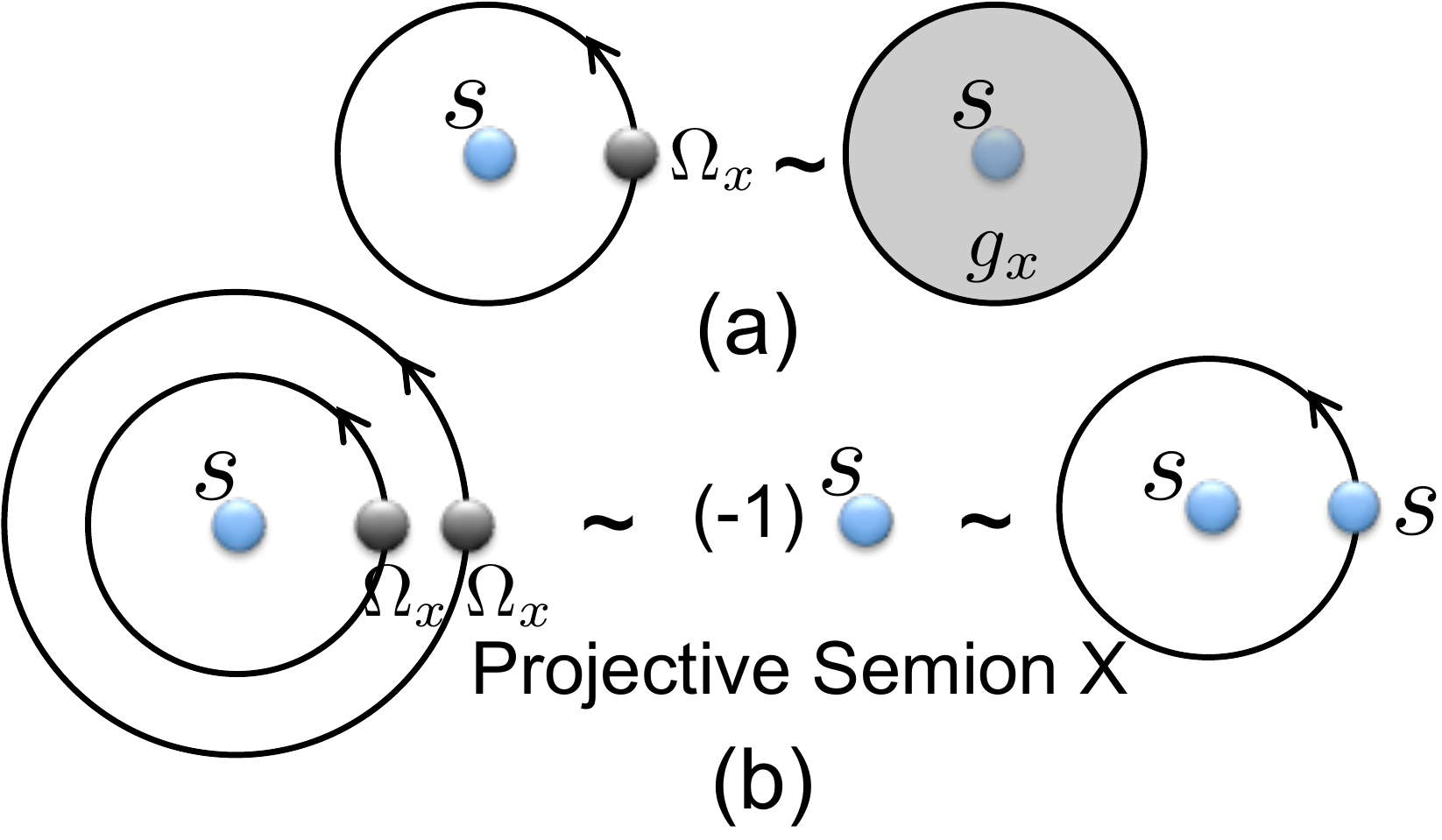}
\caption{The fusion rule $\Om_x \times \Om_x = s$ derived from the symmetry action on the semion in theory X (defined in equation \ref{ps}): (a) Bringing a gauge flux $\Om_x$ around the center semion is equivalent to acting locally on it with the corresponding symmetry $g_x$; (b) Bringing two $\Om_x$ gauge fluxes around the center semion gives rise to a $-1$ phase factor, because in theory X a semion carries half of the corresponding $g_x$ charge.  This $-1$ can be reproduced by bringing another semion around the center one, giving rise to the $s$ on the right hand side of the fusion rule.}
\label{OmOm}
\end{center}
\end{figure}

These `projective' fusion rules can be determined from the action of the symmetry on the semion. Consider for example the anomalous projective semion X state.  One important observation is that bringing a gauge flux around the semion is equivalent to locally acting with the corresponding symmetry on it, as shown in Fig.\ref{OmOm} (a).  Because the semion carries a half charge of $g_x$, bringing two $\Om_x$'s around it gives rise to a $-1$ phase factor, which can be reproduced by braiding an auxilliary semion around it.
Therefore, if we imagine fusing the two $\Om_x$ fluxes before bringing them around the semion - a distinction which should not change the global phase factor - we are led to the conclusion that two $\Om_x$'s fuse into a semion
\be
\Om_x \times \Om_x = s
\ee
Similarly, we find
\be \label{VortFuse}
\begin{array}{lll}
\Om_y \times \Om_y = I & \Om_z \times \Om_z = I & \Om_x \times \Om_y = s \Om_z \\
\Om_y \times \Om_x = \Om_z & \Om_y \times \Om_z = \Om_x & \Om_z \times \Om_y = s \Om_x \\
\Om_z \times \Om_x = s \Om_y & \Om_x \times \Om_z = \Om_y
\end{array}
\ee
The fusion rules involving the $s\Om$ fluxes can be correspondingly obtained by adding $s$ to both sides. For example, $s\Om_x \times \Om_x = I$.  
 Though the distinction between $\Om_x$ and $s \Om_x$ is arbitrary, the distinction between $\Om_x \times \Om_x$ and $s \Om_x \times \Om_x$ is not, since the two act differently on the semion.

In this way, we have established a `projective' fusion rule for the $\mathbb{Z}_2 \times \mathbb{Z}_2$ gauge fluxes in the theory X. The fusion rule is `projective' in the sense that it only obeys the $\mathbb{Z}_2 \times \mathbb{Z}_2$ group relations up to a semion.  It can be compactly expressed as a mapping $\omega$ from two group elements to the set $\{1,s\}$: $\Om_g \times \Om_h = \omega(g,h) \Om_{gh} \label{OmxOm}$.  Explicitly, for the theory X:
\be
\begin{array}{lll}
\omega(g_x,g_x) = s & \omega(g_y,g_y) = I & \omega(g_z,g_z) = I \\
\omega(g_x,g_y) = s & \omega(g_y,g_x) = I & \omega(g_y,g_z) = I \\
\omega(g_z,g_y) = s & \omega(g_z,g_x) = s & \omega(g_x,g_z) = I
\end{array}
\label{om}
\ee

The mapping $\omega$ obeys certain relations.  First, consider the fusion of three gauge fluxes $\Om_g$, $\Om_h$ and $\Om_k$. The result should not depend on which pair we fuse first: we can choose to fuse $\Om_g$, $\Om_h$ together first and then fuse with $\Om_k$ or we can choose to fuse $\Om_h$, $\Om_k$ together first and then fuse with $\Om_g$. The equivalence between these two procedures leads to the following relation among the semion coefficients:
\be
\omega(g,h) \omega(gh,k) = \omega(h,k) \omega(g,hk)
\label{H2om}
\ee
It is straight forward to check that this relation is satisfied by the $\omega$ given in Eq. \ref{om} and we are going to use this relation in our discussion of the next section.

For the other two projective semion states, similar fusion rules can be derived, also with coefficients taking semion values. In fact, in an SET phase (anomalous or not) with discrete unitary symmetries which do not change anyon types, it is generally true that when the symmetry is gauged the (putative) gauge fluxes must satisfy a projective fusion rule with coefficients in the abelian anyons.  We will discuss the general situation - including the meaning of gauge fluxes in anomalous theories - in section \ref{disc}.

Note that the fusion product of two $\Om$'s is order dependent. For example, $\Om_x \times \Om_y = s \Om_z$ while $\Om_y \times \Om_x = \Om_z$, which is very different from the usual fusion rules we see in a topological theory.  This is because the $\Om$'s discussed here are not really quasi-particles, but rather the end points of symmetry defect lines introduced in the original SET. Because the $\Om$'s are all attached to defect lines, they are actually confined. We can still define fusion between them, but because of the existence of defect lines their fusion need not be commutative. If the SET is not anomalous, the gauged theory is obtained in two steps: first the (confined) $\Om$'s should form what is called a `fusion category', whose properties are discussed in for example Ref. \onlinecite{Etingof2005,Fidkowski2014}. The fusion product between objects in a fusion category can be non-commutative, but it does have to be associative, and the pentagon relation\cite{Kitaev2006}, illustrated in Fig. \ref{pen} must still be satisfied.  If the $\Om$'s form a valid fusion category (as occurs for the CSL), the SET is non-anomalous.  In this case a gauge field can be introduced and the $\Om$'s can be promoted to deconfined quasi-particles. However, as we shall see in the following, the $\Om$'s in the anomalous projective semion theories do not even admit consistent fusion rules which satisfy the pentagon equation, and cannot be promoted to real quasi-particles with extra braiding structure.\cite{Note2}

It is worth emphasizing at this point that a projective fusion rule for the fluxes is not in itself an indication of a surface anomaly.  Indeed, a projective fusion rule for the gauge fluxes exists in many SETs, in particular the CSL.  Also, if we considered the toric code instead of the semion topological order, we could make the spinon or vison carry the sort of fractional $\mathbb{Z}_2 \times \mathbb{Z}_2$ charges that make the semion theory anomalous, but not have any problem realizing it in a $\mathbb{Z}_2 \times \mathbb{Z}_2$ symmetric way in $2D$.  To expose the anomaly in the projective semion model, we have to do more work.

\subsection{Anomaly in the statistics of gauge fluxes}

In this subsection we derive the anomaly using only the statistics of the anyons and gauge fluxes.  The argument given here is a physical interpretation of the obstruction calculation in section 8.6 of Ref. \onlinecite{Etingof2010}.

First, recall the information that goes into defining the anyon topological order: it is encoded in two sets of data, the statistics of exchanging anyon $a$ with anyon $b$ (`R-matrix') and the fusion `statistics' in the associativity of fusing anyons $a$, $b$ and $c$ (`F-matrix'). In an abelian theory, the fusion statistics are the phase differences between the following two processes: one which first fuses $a$ with $b$ and then fuses the product with $c$ (i.e. $(a \times b) \times c$) and the one which first fuses $b$, $c$, and then fuses $a$ with the result (i.e. $a \times (b \times c)$) (in non-abelian theories they are isomorphisms between direct sums of tensor products of the corresponding fusion spaces).

\begin{figure}[htbp]
\begin{center}
\includegraphics[width=8.0cm]{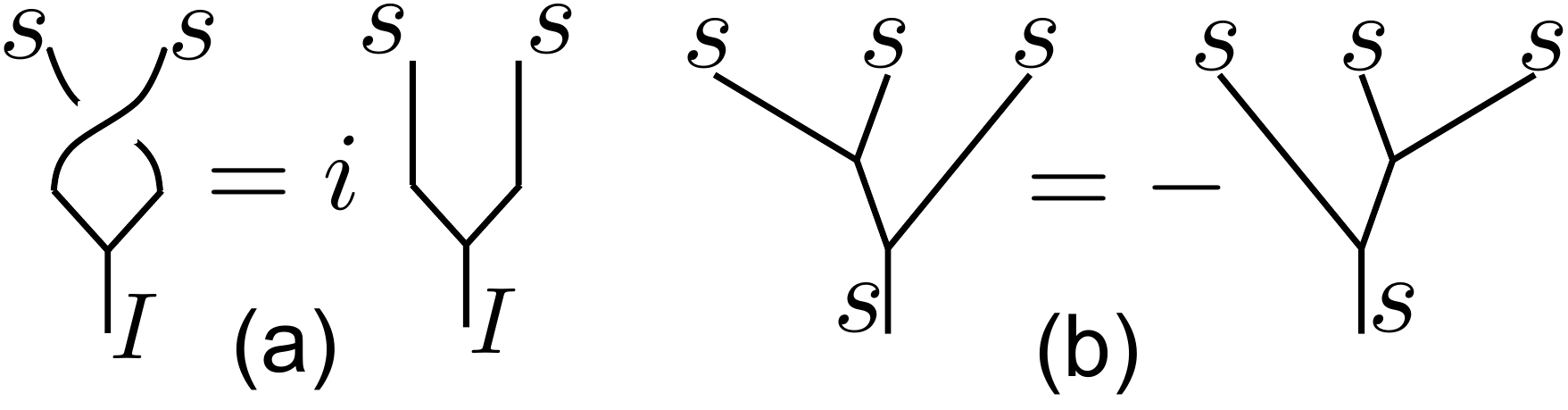}
\caption{Braiding and fusion statistics of the semion theory: (a) The exchange of two semions leads to a phase factor of $i$; (b) The two ways of fusing three semions differ by a sign.}
\label{RFs}
\end{center}
\end{figure}

For example, in the semion theory discussed here, the only braiding statistics happen when two semions are exchanged, as shown diagrammatically in Fig. \ref{RFs} (a). The exchange of two semions leads to a phase factor of $i$. Denoting the exchange statistics by $R_{\omega,\omega'}$,
we have:
\be
R_{s,s}=i, R_{\omega,\omega'}= 1 \ \text{otherwise}
\ee
The only nontrivial fusion statistics occur when three semions are fused together in different orders, as shown in Fig. \ref{RFs} (b). The two orders of fusion differ by a phase factor of $-1$. Denoting the fusion statistics as $F_{\omega,\omega',\omega''}$, we have:
\be
F_{s,s,s}=-1, F_{\omega,\omega',\omega''} = 1 \ \text{otherwise}
\ee

\begin{figure*}[htbp]
\begin{center}
\includegraphics[width=15.0cm]{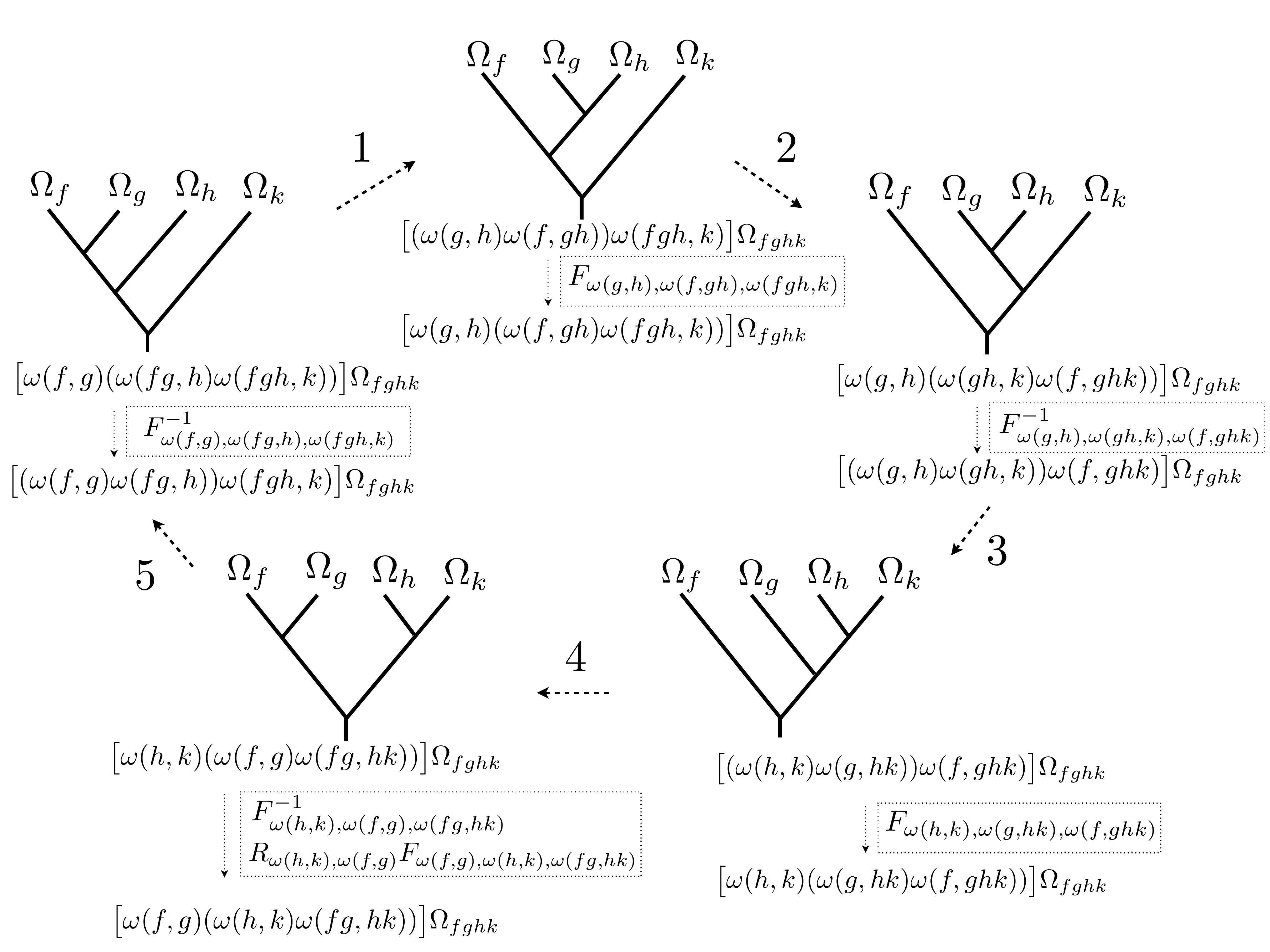}
\caption{Pentagon equation for the fusion of four gauge fluxes.}
\label{pen}
\end{center}
\end{figure*}

Now, if the symmetry in the projective semion theory can be consistently gauged, then we should similarly be able to define for the gauge fluxes not only the projective fusion rules but also the braiding and fusion statistics involved with exchanging two fluxes or fusing three of them in different orders. These statistics cannot be chosen arbitrarily, but have to satisfy certain consistency conditions, one of which is called the pentagon equation, shown in Fig. \ref{pen}. 

The pentagon equation relates different orders of fusing four gauge fluxes, $\Om_f$, $\Om_g$, $\Om_h$, and $\Om_k$. For example, in the configuration on the top left corner of Fig. \ref{pen}, $\Om_f$ and $\Om_g$ are fused together first, then with $\Om_h$ and finally with $\Om_k$. In moving from this configuration to the next configuration through step 1, we have changed the associativity in the fusion of the first three fluxes.  Such a step is related to a phase factor given by the fusion statistics of the first three gauge fluxes, analogous to the phase factor shown for three semions in Fig. \ref{RFs} (b).  The pentagon equation then states that upon traversing all 5 steps and ending up at the original configuration, the total phase factor gained should be equal to $1$.  This is simply because each configuration represents a choice of basis in the space of local operators that fuse the four fluxes to $\Om_{fghk}$, and each step is a change of basis.

For the gauge fluxes, there is a direct way to check whether this pentagon equation can be satisfied by using the projective fusion rules of the fluxes.  The key point, explained more carefully in the following subsection, is that the dependence of the three flux F-matrix $F_{f,g,h}$ on the topological superselection sector of each flux (i.e. on $\Om_f$ versus $s\Om_f$, etc.) is uniquely determined (modulo gauge equivalence) by consistency requirements once the projective fusion rules for the fluxes have been chosen, up to an overall phase factor $\beta(f,g,h)$.  We now show that, for the choice $\beta(f,g,h)=1$, the four flux pentagon equation is violated; in the next subsection we will see that this violation -- which turns out to be a phase factor $\nu(f,g,h,k)$ -- cannot be repaired simultaneously for all choices of $f,g,h,k$ by {\emph{any}} choice of $\beta(f,g,h)$.

First of all, by repeated use of the projective fusion rule, the fusion results of each of the five configurations in figure \ref{pen}, starting from the top left one, can be reduced to:

\be
\begin{array}{l}
((\omega(f,g)\omega(fg,h))\omega(fgh,k))\Om_{fghk} \\
((\omega(g,h)\omega(f,gh))\omega(fgh,k))\Om_{fghk} \\
((\omega(g,h)\omega(gh,k))\omega(f,ghk))\Om_{fghk} \\
((\omega(h,k)\omega(g,hk))\omega(f,ghk))\Om_{fghk} \\
((\omega(f,g)\omega(h,k))\omega(fg,hk))\Om_{fghk}
\end{array}
\label{pen_red}
\ee

Comparing these, we can see that the only difference between the configurations arises because of the different anyon coefficients (i.e. the product of $\omega$'s in front of $\Om_{fghk}$), whose braiding and fusion statistics are already known. Therefore, we can derive all of the associated phase factors.  Indeed, they all come from operations done on the fusion product of the three $\omega$'s done in between the five hexagon steps, as illustrated in figure \ref{pen}.  These are all just F-matrix moves which change order in which the $\omega$'s are fused, except for one special case: between step 4 and step 5 we also have an R-matrix move that exchanges $\omega(h,k)$ and $\omega(f,g)$.  The total phase acquired is the product of all the factors in the dotted boxes in figure \ref{pen}.  One can check directly that this phase factor is not equal to $1$ for some choices of the fluxes in the anomalous projective semion X ,Y, and Z theories.

Of course, as discussed above, we still have the freedom to deform the flux F-matrices by arbitrary phases that depend only on the group elements: for example, the first step would gain a phase factor $\beta(f,g,h)$, the second $\beta(f,gh,k)$, etc.  The fact that there is no choice of $\beta$ that simultaneously gauges away all the pentagon equation phases is explained in the next subsection; it reflects the fact that these pentagon phases form a non-trivial $U(1)$ valued 4-cocycle of $G$.

\subsection{Group cohomology structure of the anomaly}

In order to explain the last point above, we will make use of group cohomology theory, reviewed in appendix \ref{Gcoh}.  The discussion in this section is somewhat formal and mathematical, but given the classification of SPT phases with group cohomology, it allows us to make a direct connection to 3D SPTs. This mathematical structure was explored in the context of group extensions of fusion categories by Etingof et. al.\cite{Etingof2010} and we briefly explain the main ideas in this section. This general structure applies not only to the anomalous projective semion example, but to all SETs with discrete unitary symmetries.

First, Eq. \ref{H2om} implies that the gauge fluxes form a projective representation of the symmetry group $\mathbb{Z}_2 \times \mathbb{Z}_2$, with semion-valued coefficients.  Because the semion has a $\mathbb{Z}_2$ fusion rule, namely $s\times s=1$, the projective fusion rules of the gauge fluxes can be classified by $H^2(\mathbb{Z}_2 \times \mathbb{Z}_2, \mathbb{Z}_2)$. Direct calculation shows that $H^2(\mathbb{Z}_2 \times \mathbb{Z}_2, \mathbb{Z}_2) = \mathbb{Z}_2 \times \mathbb{Z}_2 \times \mathbb{Z}_2$.  One of the three $\mathbb{Z}_2$'s on the right side of this equation determines whether or not the semion is a spin $1/2$; among the theories where it is a spin $1/2$, the remaining $\mathbb{Z}_2 \times \mathbb{Z}_2$ parametrizes the CSL state and the three anomalous projective semion (X,Y,Z) states (among the theories where the semion is not a spin $1/2$, the remaining $\mathbb{Z}_2 \times \mathbb{Z}_2$ parametrizes non-anomalous theories where this semion can carry half-integral charges of either ${\mathbb Z}_2$).

It is generally true that possible projective fusion rules, hence possible symmetry fractionalization patterns, in an SET theory are classified by $H^2(G,A)$ where $G$ is the symmetry group and $A$ denotes the (abelian fusion group of) abelian anyons\cite{Essin2013,Fidkowski2014}. Notice that here the coefficients of the projective fusion rule are valued only in the abelian sector of the topological theory.  In more general situations, the anyons can be permuted by the symmetry in the system. Correspondingly, the $G$ action on the coefficients $A$ can be nontrivial, but the $H^2(G,A)$ classification as discussed in appendix \ref{Gcoh} turns out to still apply \cite{Fidkowski2014, Maissam2014}.  

With the symmetry fractionalization information encoded in $\om \in H^2(G,A)$, we can then move on to determine whether the SET theory is anomalous or not, i.e. do we get a consistent extended topological theory involving both the gauge charges, gauge fluxes and the original anyons, upon gauging the $G$ symmetry?  This is a highly nontrivial process, but fortuitously a mathematical framework has been developed in Ref. \onlinecite{Etingof2010}, concerning group extensions of braided categories, which is precisely equipped to deal with this issue.  Specifically, this mathematical framework allows us to draw the following important conclusions:

1. In order to determine whether an SET theory is anomalous or not, we only need to look at the fusion and braiding statistics of the original anyons and all the gauge fluxes. If a consistent topological theory can be defined for the original anyons and all the gauge fluxes, then gauge charges can always be incorporated without obstruction.

Therefore, to detect an anomaly in a putative SET theory, we need to look for possible ways to define consistent fusion and braiding statistics for all the gauge fluxes $\Om$ and the original anyons $\al$. We already know the fusion and braiding statistics of $\al$ - they are given by $R_{\al,\al'}$ and $F_{\al,\al',\al''}$ (for simplicity of notation we use the same label for the set of anyons in $\al$ and $\Om$. Also we ignore the complication in notation due to nonabelian anyons as long as it does not cause confusion).  What we now need to determine are the fusion and braiding statistics of the $\Om$ with $\al$, and the $\Om$ with $\Om$: $R_{\al,\Om}$, $R_{\Om,\Om'}$, $F_{\al,\al',\Om}$, $F_{\al,\Om,\Om'}$ and $F_{\Om,\Om',\Om''}$, and all permutations of these indices.

The strategy is to bootstrap from the known data ($R_{\al,\al}$ and $F_{\al,\al,\al}$) and solve for the unknowns using the consistency equations they have to satisfy. The consistency equations come in two types: pentagon equation involving the fusion statistics $F$ and hexagon equation involving both the braiding and the fusion statistics $R$ and $F$. (For a detailed discussion of these equations see Ref. \onlinecite{Kitaev2006}.) There is one pentagon equation for every combination of 4 quasiparticles (including both the $\al$ and $\Om$) and there are two hexagon equations for 3 quasiparticles (including both the $\al$ and $\Om$). Of course the pentagon and hexagon equations involving only $\al$ are all satisfied. The next step is to include one $\Om$ and try to solve for the $R_{\al,\Om}$ and $F_{\al,\al,\Om}$ that appear in those equations and so on. If the equations have no solutions, then we have detected an anomaly. Of course, this would be a lengthy process to follow by brute force. Luckily, there are only two steps at which the equations might have no solution: \cite{Etingof2010,Fidkowski2014, Maissam2014}

2.  The first such step results in an obstruction in $H^3(G,A)$, which, when non-zero, signals a lack of any solutions.  However, this obstruction appears only when the group $G$ acts non-trivially on the anyons by permutation (technically, this is only true for an abelian theory; in the general case the data is the action by braided autoequivalence \cite{Etingof2010, Fidkowski2014, Maissam2014}), and essentially corresponds to the fact that, when it is non-zero, it is impossible to define an associative fusion product for the fluxes (nevermind requiring the associativity constraint to satisfy the pentagon equation).  Since we are only concerned in this paper with a trivial permutation by $G$ on the set of anyons,  we can safely ignore this obstruction.

3. When the first type of obstruction vanishes, there is still a possibility that we run into a second type of obstruction when we try to satisfy the pentagon equation of four $\Om$'s, as is the case with the projective semion example.  From figure \ref{pen}, the total phase factor up to which the pentagon equation is satisfied is:
\be
\begin{array}{l}
\nu(f,g,h,k) = R_{\omega(h,k),\omega(f,g)} \times  \\
F_{\omega(g,h),\omega(f,gh),\omega(fgh,k)}F^{-1}_{\omega(g,h),\omega(gh,k),\omega(f,ghk)}\\
F_{\omega(f,g),\omega(h,k),\omega(fg,hk)}F^{-1}_{\omega(f,g),\omega(fg,h),\omega(fgh,k)} \\
F_{\omega(h,k),\omega(g,hk),\omega(f,ghk)}F^{-1}_{\omega(h,k),\omega(f,g),\omega(fg,hk)}
\end{array}
\label{ob}
\ee
for $f,g,h,k \in G$. It was proved in Ref.\onlinecite{Etingof2010} that the $\nu(f,g,h,k)$ data forms a 4-cocycle of the group $G$ with $U(1)$ coefficients (where the unitary $G$ acts trivially on $U(1)$).  Furthermore, as mentioned above, there is an overall phase ambiguity in the F-matrix of three fluxes; modifying this F-matrix by the phase $\beta(f,g,h)$ leaves all consistency conditions satisfied, and modifies $\nu(f,g,h,k)$ by:
\be
\frac{\bt(f,g,h)\bt(f,gh,k)\bt(g,h,k)}{\bt(fg,h,k)\bt(f,g,hk)}
\ee
In other words, $\nu(f,g,h,k)$ has some gauge - or co-boundary -  degrees of freedom.  Therefore, this type of obstruction is classified by $H^4(G,U(1))$, which corresponds exactly to the classification of 3D SPT phases with $G$ symmetry.

The explanation in this section provides a very brief physical interpretation of some of the results in Ref. \onlinecite{Etingof2010} (see \onlinecite{Fidkowski2014} for more details).  The upshot of this section is the formula for the anomaly given in Eq. \ref{ob}. For the anomalous projective semion theories, we calculated this $\nu(f,g,h,k)$, confirmed that it is a four cocycle, and checked that it is indeed non-trivial both analytically (see appendix \ref{4cocycle}) and numerically.  We have also confirmed that for the non-anomalous chiral spin liquid, $\nu(f,g,h,k)$ is a trivial 4-cocycle.  The second main point of this section is the connection between the SET anomalies and the SPT phases. They are both classified by $H^4(G,U(1))$ and hence one might expect a close relation between the two, which we demonstrate in the next section.

\section{Realizing the anomalous projective semion models on the surface of 3D SPTs}
\label{WW}

In section \ref{anomaly}, we established that a system with excitations that have semionic statistics and transform under a ${\mathbb{Z}}_2 \times {\mathbb{Z}}_2$ global symmetry projectively according to equation \ref{ps} cannot be realized purely in 2 dimensions.  In this section, we present an exactly solvable 3+1D model that realizes the projective semion theory at its 2+1D boundary.  Our approach is based on the Walker-Wang construction \cite{Walker12}, which gives a family of exactly solved lattice models with trivial gapped bulk and surface topological order.  Our particular Hamiltonian is a decorated version of the Walker-Wang model that describes a 3D loop gas of semion strings \cite{vonkeyserlingk13}.  Essentially, we decorate this Walker-Wang model with unitary linear representations of $G$ in such a way that semion loops bind certain Haldane chains of $G=\mz_2 \times \mz_2$.  In this way, the surface semion excitation, which is just an endpoint of an open semion string, transforms in the required projective representations.  Before delving into the details of the decoration, however, we briefly review the semion Walker-Wang model itself (for a more detailed discussion readers should consult  Ref. \onlinecite{vonkeyserlingk13}).

\subsection{The Walker-Wang semion model: a brief review}

\begin{figure}[htbp]
\begin{center}
\begin{tabular}{ll}
(a)  & \\
& \includegraphics[width=6.5cm]{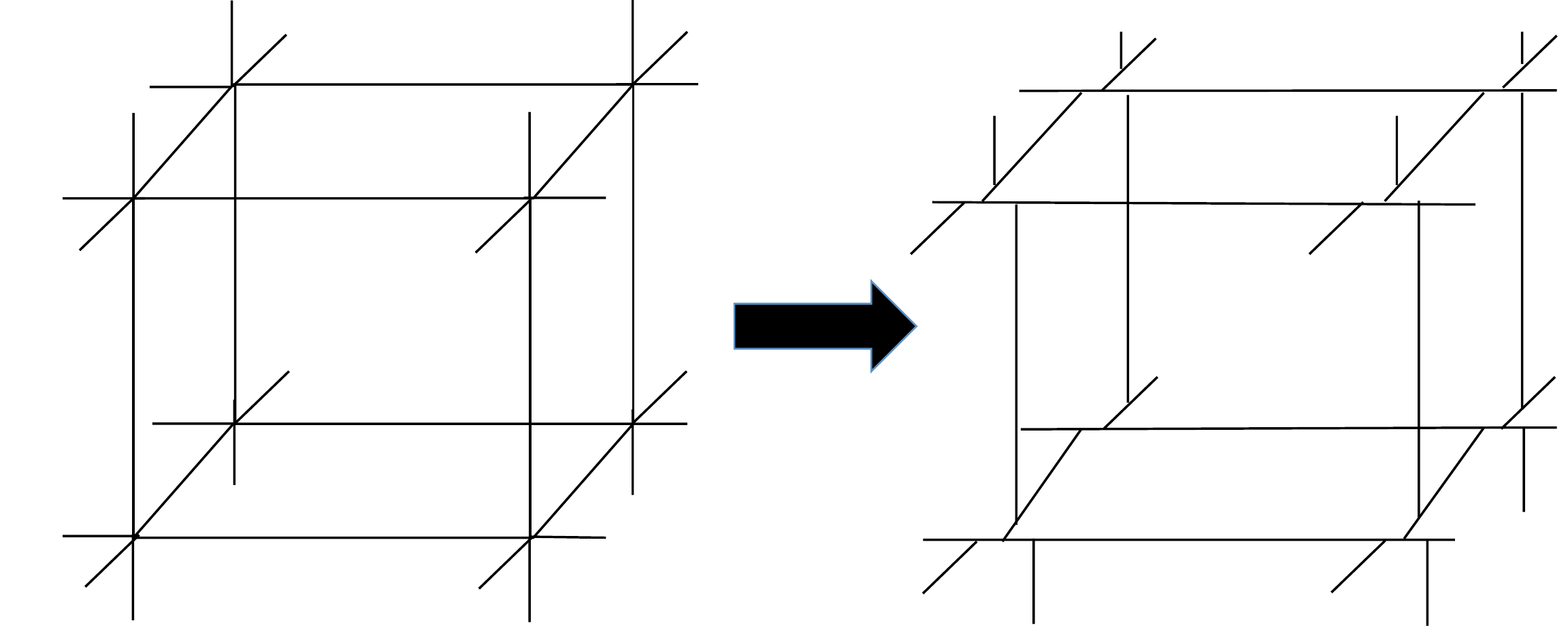} \\
(b)& \\
& \includegraphics[width=7.0cm]{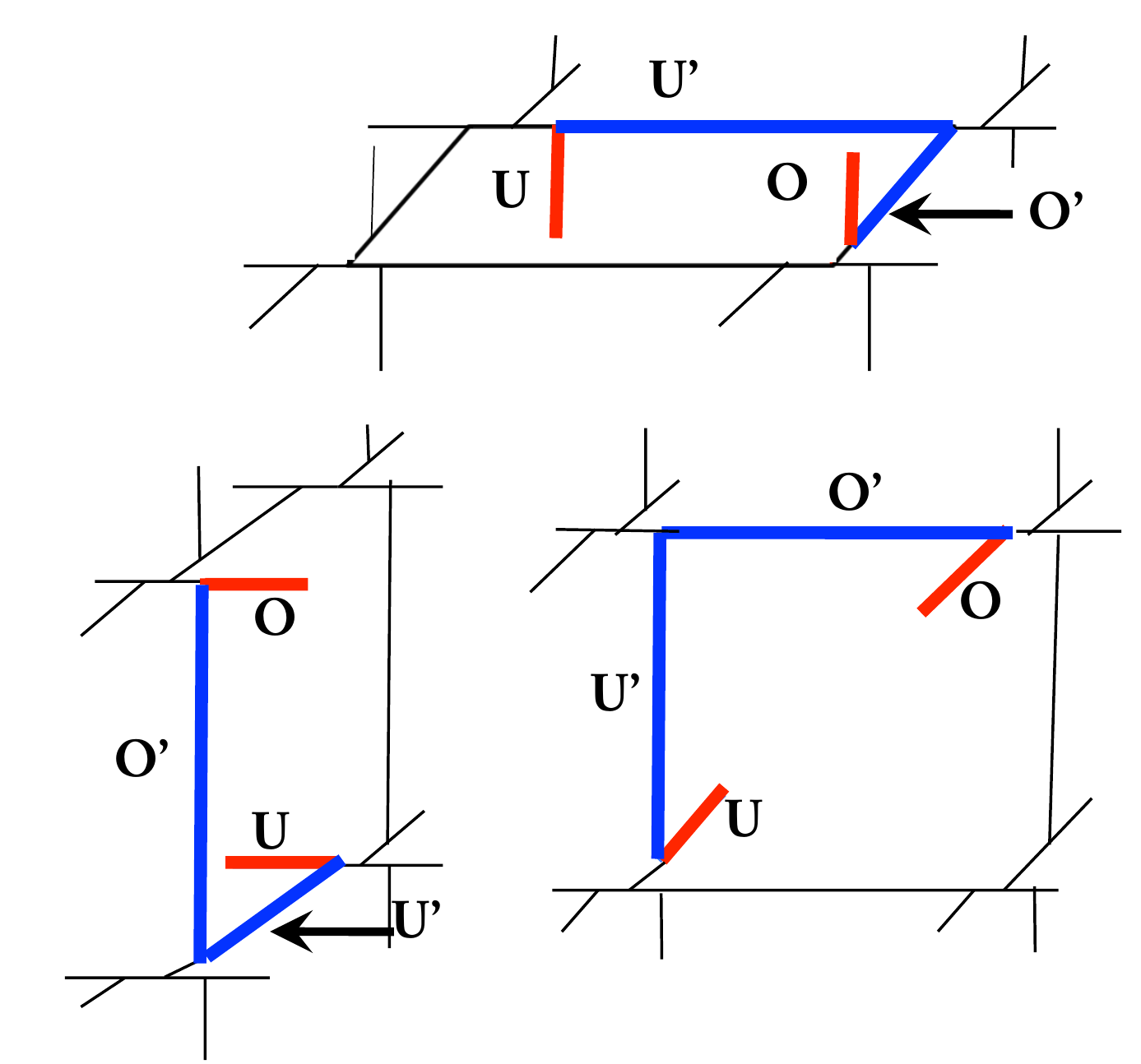} 
\end{tabular}
\caption{(a) We define our model is defined on the point-split cubic lattice, because the Hamiltonian is simplest for a lattice with only trivalent vertices.  
(b) The plaquette term is defined with respect to a fixed angle of projection.  In this angle, there are two edges ($O$ and $U$) that are projected into each plaquette from above and below, respectively, shown in red. These  (and their partner edges $O'$ and $U'$, shown here in blue) are used to define the phase factors $\Phi_{O, O'}, \Phi_{U, U'}$ in the plaquette operator.  }
\label{WWPlaqFig}
\end{center}
\end{figure}

The Walker-Wang models are most easily described on a trivalent 3d lattice, obtained by starting with a cubic lattice and ``point-splitting" each vertex into 3 trivalent vertices (Fig. \ref{WWPlaqFig} a).  The model can also be described on the cubic lattice, but in this geometry the Hamiltonian requires additional terms; here we will therefore discuss our model on the more complicated point-split cubic lattice. 

The Hilbert space consists of a hardcore boson with occupation $n_i=0,1$ on each edge of the lattice.  The Hamiltonian is a sum of two commuting terms, one acting on vertices in the lattice and one on plaquettes:
\be \label{WWHam}
H = - \sum_V A_V - \sum_P B_P
\ee

\begin{figure}[htbp]
\begin{center}
\includegraphics[width=8.0cm]{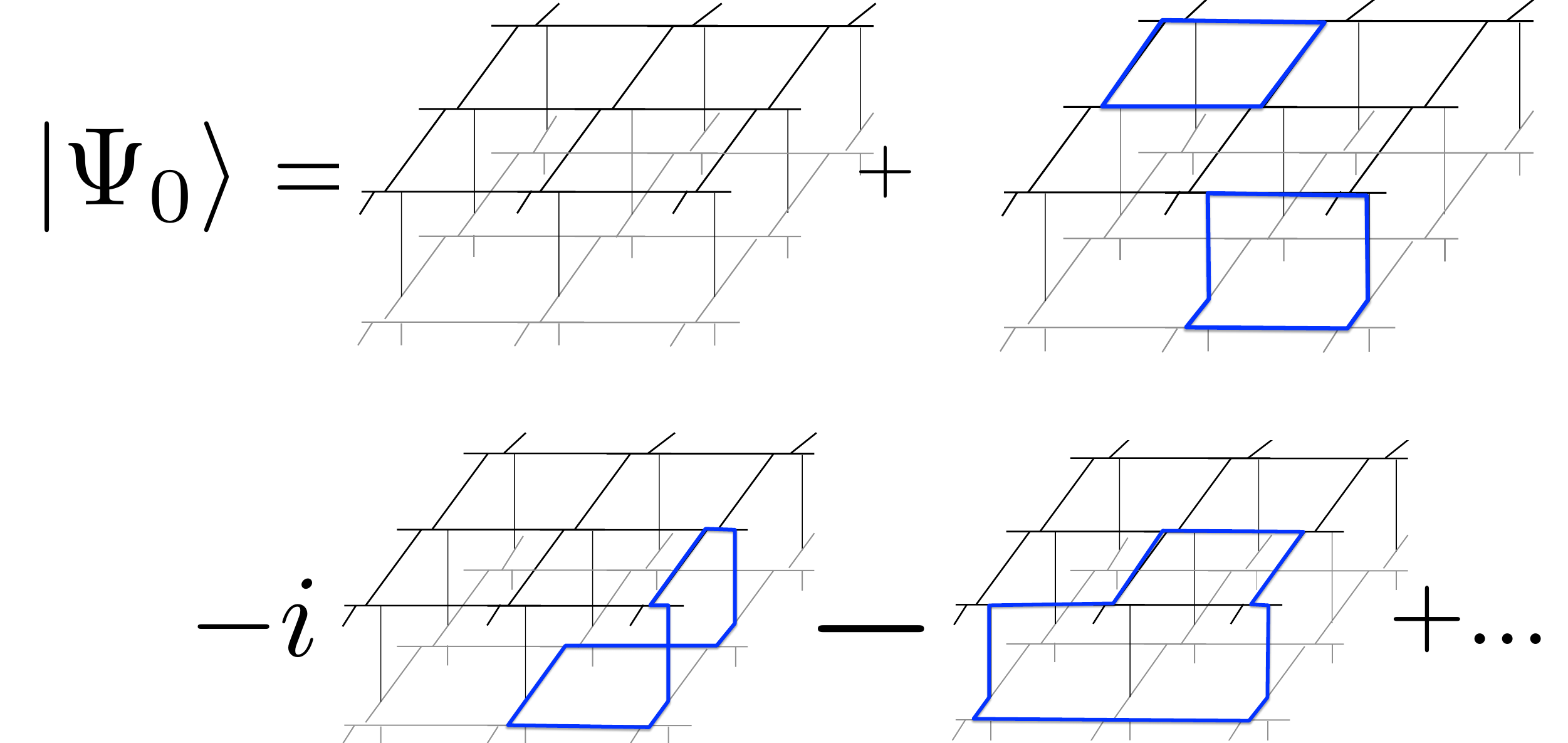}
\caption{Some configurations in the semion ground state.  The vertex condition ensures that only configurations with an even number of edges on which $n_i=1$ (shown in blue in the Figure) can meet at a vertex, such that the ground state is a superposition of loops.  The relative amplitudes of these loop confiurations are given by the phase factor $ \Theta (P) \Phi_{O,O'} \Phi_{U,U'}$ in Eq. (\ref{SemPlaqH}).}
\label{GSFig1}
\end{center}
\end{figure}

We define
\be \label{SemVert}
A_V = \prod_{i \in ^*V} \tau^z_i, 
\ee
where $\tau^z_i  = 1 - 2 n_i$ and $^*V$ is the set of 6 edges entering the vertex $V$.  These terms, all of which commute, favor configurations in which the number of spin-down edges at the vertex is even.  The plaquette terms will also turn out to commute with the vertex terms (and each other), so that we can think of the model as a loop gas (Fig. \ref{GSFig1}): edges on which $n_i=1$ (blue in the figure) form closed loops if $A_V=1$ at each vertex.  

$B_P$ is defined to simultaneously change occupation numbers $n_i$ on all edges $i$ around a plaquette and assign a configuration-dependent phase to the result.  Specifically, introducing the operator $\tau^x_i=|0\rangle_i\langle1|_i+|1\rangle_i\langle 0|_i$, we have:
\be \label{SemPlaqH}
B_P = -  C(P) \left (\prod_{i\in \partial P} \tau^x_i  \right ) \Theta (P) \Phi_{O,O'} \Phi_{U,U'},
\ee
where  $\partial P$ is the set of all edges bordering $P$.  Here $C(P)$ is defined to be equal to $1$ only for configurations in which satisfy the vertex terms at $P$, and $0$ otherwise:
\be
C(P)=\prod_{V\in P} ((1+A_V)/2)
\ee
As in the 3D Toric code\cite{Hamma05,CastelnovoChamon}, the product over $\tau^x$ ensures that the ground state is a superposition of closed loops.  Their relative coefficients are determined by the rest of the expression on the right hand side of Eq. (\ref{SemPlaqH}), which will be a phase factor that depends only on $\{n_i\}$ on edges on or adjacent to the plaquette $P$.   

Before defining the phase factors $\Theta (P),  \Phi_{O,O'}, \Phi_{U,U'}$, we remark that they will be chosen in such a way that the coefficients of different loop configurations in the ground state of the loop gas reproduce the Euclidean partition function $Z^{CS}_{ \{L \}}$ of a 3D $U(1)_2$ Chern-Simons theory (a ``semion loop gas").  To make this connection, one needs to interpret the Walker-Wang model as a particular lattice discretization of a theory of framed loops; every linking of loops then incurs a factor of $-1$, while every counter-clockwise (clockwise) 360 degree twist in a loop induces a factor of $i$ ($-i$), and furthermore there is also a phase factor of $(-1)^{N_L}$, where $N_L$ is the number of loops.  We refer the interested reader to references \onlinecite{Walker12, vonkeyserlingk13} for more details.

Returning to the definition of our plaquette term, we let
\be
\Theta (P) =  \left (\prod_{j\in ^*P} i^{n_j}  \right ) 
\ee
where $^*P$ denotes the set of all edges entering the plaquette $P$.  The role of $\Theta(P)$ is in fact to reproduce the factor of $(-1)^{N_L}$ mentioned above; indeed, $\Theta(P)$ ensures that if $\prod_{i \in \partial P} \tau^x_i$ changes the number of loops in a particular loop configuration, then $-\Theta(P)=-1$ for that configuration.

Furthermore, we define
\be \label{CSPhase}
 \Phi_{O, O'}=   i^{n_O ( 1 + 2 n_{O'}) } \ , \ \ \   \Phi_{U,U'} = i^{ n_U ( 1 + 2 n_{U'}) }
\ee
This definition is represented graphically in Fig. \ref{WWPlaqFig}, and involves a particular 2d projection of the cubic lattice.  Using this projection, for each plaquette we can define a set of special edges $O$ (for ``over") and $U$ (for ``under") that project into the plaquette from above and below, respectively.  $O'$ and $U'$ are neighbouring edges in $\partial P$, as shown in Fig. \ref{WWPlaqFig}.  Again, if we think of our model as a lattice discretization of the theory of framed loops mentioned above, the role of $\Phi_{O, O'}$ and $\Phi_{U,U'}$ is to reproduce the factors of $\pm i$ due to twisting and $-1$ due to linking loops.

Having defined the Hamiltonian, let us now analyze it.  First of all, it is immediately clear that $[ A_{V}, A_{V'} ] = [ B_{P}, A_{V} ] = 0$ for all $V,V',P$.  Though not as immediately apparent, it is also true that $[ B_{P}, B_{P'} ] =0$ for all pairs of plaquettes $P,P'$ (to check this, note that the only non-trivial case is that of edge-sharing $P$ and $P'$ intersecting at a 90 degree angle).  Thus the ground state is a simultaneous eigenstate of all $A_V, B_P$.  Furthermore, these terms are unfrustrated, i.e. there exists a ground state with all $A_V,B_P=1$; see e.g. reference \onlinecite{Walker12}.  This ground state is precisely the semion loop gas described above.  

Having understood the ground state, let us study the excitations.  One may imagine that, as in the toric code, the spectrum contains both closed loop excitations for which $B_P = -1$, and point excitations $A_V=-1$ which are violations of vertex terms.  However, because of the phases introduced in the plaquette term above, this guess is incorrect: putative string operators that create such excitations always violate a line of plaquettes, and all point-like excitations have a linear confining energy cost \cite{vonkeyserlingk13}.

\begin{figure}[htbp]
\begin{center}
\includegraphics[width=7.0cm]{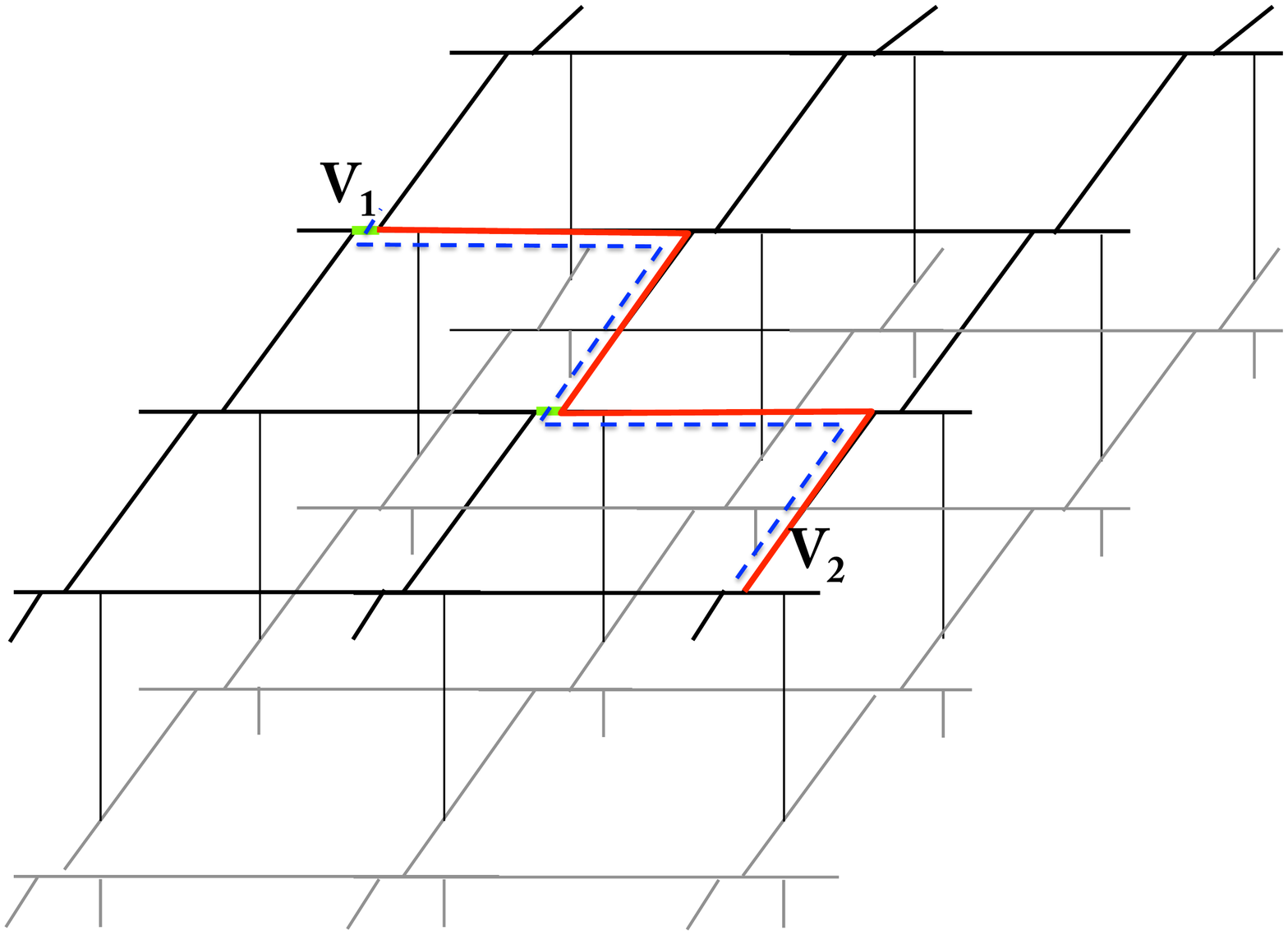}
\caption{The semion string operator on the surface.  The curve $C$ connecting the surface vertices $V_1$ and $V_2$ is shown in red.  A second copy of $C$ (shown in dashed blue), displaced from $C$ along the $(-1,-1,1)$ direction, determines the set of $R$-vertices.  These are vertices where the dashed blue curve crosses over an edge $e_i$ (shown in green) of the surface.  The edge $i+1$ is the edge in $C$ that shares a vertex with $e_i$ and is crossed second when following $C$ such that $C$ turns to the right at these crossings.  }
\label{SemSurfFig}
\end{center}
\end{figure}

\begin{figure}[htbp]
\begin{center}
\includegraphics[width=7.0cm]{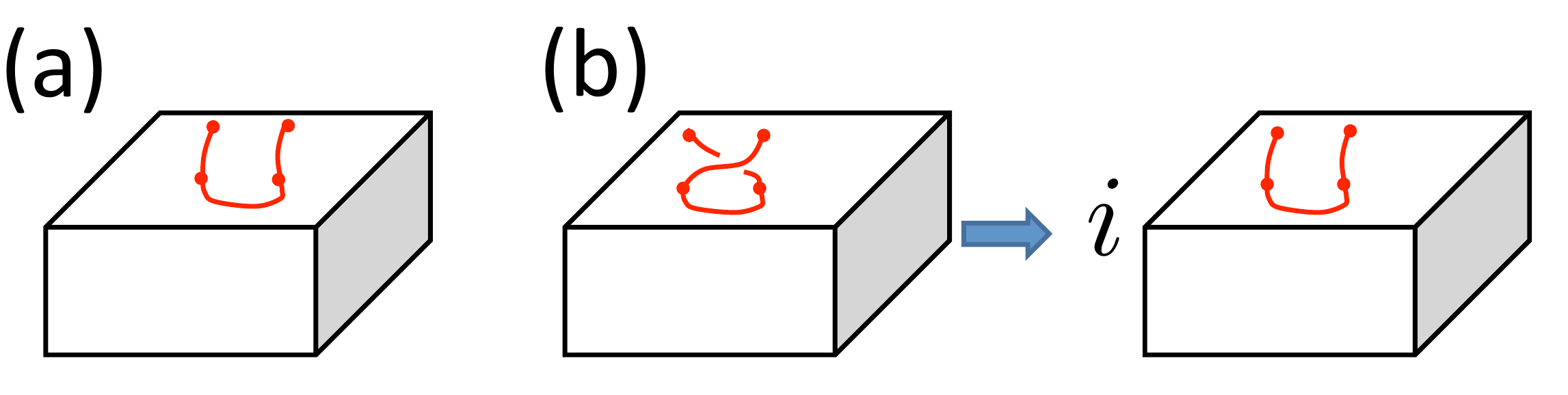}
\caption{Beginning with the configuration shown in (a), we exchanging two open string ends at the surface.  Returning the strings to their original configuration (b) multiplies the eigenstate of the Walker-Wang model by $\pm i$, demonstrating that these defects have semionic statistics.}
\label{ExchangeFig}
\end{center}
\end{figure}

On the surface, however, vertex defects are deconfined: if $V_1$ and $V_2$ are vertices on the boundary of our 3D system, there exist surface string operators that create states in which the eigenvalues obey $A_{V_1} = A_{V_2} =-1$, but $B_P =1$ everywhere.  The corresponding excited states will consist of superpositions of closed loops, together with an open string connecting the vertices $V_1$ and $V_2$.  They are created by the semion string operator
\be \label{SString}
\hat{S}_{C} = \prod_{i \in C} \tilde{\tau}^x (-1)^{n_i ( 1- n_{i+1})} \prod_{R-vertices} i^{n_{e_i} + 2 n_{i+1}}
\ee
where $C$ is any path connecting $V_1$ and $V_2$ (see Fig. \ref{SemSurfFig}).   These vertex defects are semions: exchanging them multiplies the excited-state wave function by a phase factor of $\pm i$ (Fig. \ref{ExchangeFig}).  Both the bulk and surface spectrum are derived in detail in Ref. \onlinecite{vonkeyserlingk13}.

\subsection{Decorating the model with a global symmetry}

We next show how to decorate this model such that the surface semion transforms projectively under a global $\mathbb{Z}_2 \times \mathbb{Z}_2$ symmetry.  Here we discuss a physically intuitive decoration, which will be useful in understanding the anomalous theory with a gobal symmetry.    In section \ref{3loopbraiding} we will introduce a more abstract decorated model, which behaves identically to the model discussed here when the symmetry is global but is more straightforwardly gauged; this  will be useful to compute the three loop braiding statistics.

We begin by enlarging our Hilbert space by adding three auxilliary degrees of freedom at each vertex (one associated with each edge).  
For the sake of illustration, we first consider case of the non-anomalous chiral spin liquid (CSL) surface, corresponding to a trivial SPT, in which each auxillary degree of freedom can be either a spin $1/2$ (transforming projectively under $\mathbb{Z}_2\times \mathbb{Z}_2 $) or a spinless particle $b$ (transforming in the singlet representation of $\mathbb{Z}_2\times \mathbb{Z}_2$).  (We will describe how to modify this construction to obtain a projective semion model at the end of this section).  
We further impose the constraint that, within this enlarged Hilbert space, only states with an even number of spin $1/2$'s at each vertex are allowed.  This ensures that the vertex degree of freedom transforms in a linear representation of $\mathbb{Z}_2\times \mathbb{Z}_2$.

Within this enlarged Hilbert space, it is convenient to define the modified edge variables
\be \label{Constr01}
\ket{\tilde{0}}=\ket{0}\ket{b_i b_j } 
\ , \ \  
\ket{\tilde{1}}=\ket{1}\left (\ket{\uparrow_i \downarrow_j } - \ket{ \downarrow_i \uparrow_j } \right )/ \sqrt{2}
\ee
where $\uparrow, \downarrow$ represent the physical spin states, $b$ are the spinless states, and $i,j$ index the spin variables associated with the two ends of the edge.  (We refer to $|\uparrow \rangle, |\downarrow \rangle,$ and $|b \rangle$ collectively as auxiliary states).  $\ket{\tilde{1}}$ describes an edge in which a hard-core boson -- representing semion occupation on the edge -- occurs in conjunction with two spin degrees of freedom, which combine to form a singlet.  $\ket{\tilde{0}}$ is an edge with no boson -- \emph{i.e.} no semion -- and where the auxiliary degrees of freedom are both spinless.  

We can express our Hamiltonian in terms of the decorated spin operators:
 \be
 \tilde{\tau}^z_i =  \ket{\tilde{0}_i}\bra{\tilde{0}_i}-\ket{\tilde{1}_i}\bra{\tilde{1}_i} \ , \ \ \ 
 \tilde{\tau}^x_i =  \ket{\tilde{0}_i}\bra{\tilde{1}_i}+\ket{\tilde{1}_i}\bra{\tilde{0}_i}
\ee
These act like Pauli spin operators within the Hilbert space (\ref{Constr01}): by construction, $( \tilde{\tau}^x)^2 =( \tilde{\tau}^z)^2 =1,$ and $ \tilde{\tau}^x \tilde{\tau}^z =-  \tilde{\tau}^z  \tilde{\tau}^x$.  Importantly, they act only on the set of auxiliary vertex variables associated with the edge in question, so that decorated spin operators acting on adjacent edges commute.  Note also that $\tilde{\tau}^x$ conserves the total spin, since it replaces a singlet-bonded pair of spin-1/2's at adjacent vertices with a pair of spin 0 particles.  

The Hamiltonian is:
\be
H = H_0 -\sum_V \tilde{A}_V -\sum_P \tilde{B}_P
\ee
where $ \tilde{A}_V, \tilde{B}_P$ are as in Eq.'s (\ref{SemPlaqH}, \ref{SemVert}), with $\tau^{x,z}_i$ replaced by $\tilde{\tau}^{x,z}_i$, and $n_i$ by $\tilde{n}_i = \frac{1}{2} \left( 1 - \tilde{\tau}^z_i \right)$.  $H_0$ is a potential term favoring the edge states $ \ket{\tilde{0}}$ and $\ket{\tilde{1}}$: 
\be
H_0=-\ket{\tilde{0}}\bra{\tilde{0}}-\ket{\tilde{1}}\bra{\tilde{1}}
\ee

\begin{figure}[h]
  \centering
     \includegraphics[width=3.4in]{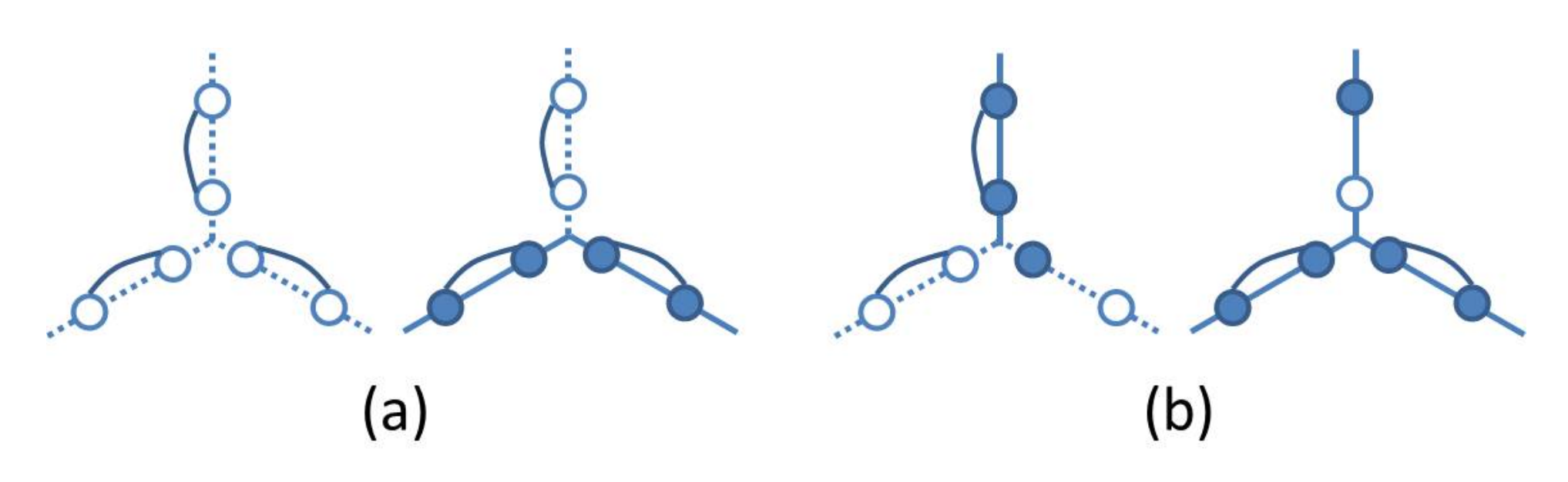}
	\caption{(a) low energy and (b) high energy configurations at each vertex. Dotted lines represent links without strings while solid lines represent links with strings. Empty circles represent spinless vertex particles while solid circles represent spin-1/2's. Two connected circles form a singlet.}
	\label{dec}
\end{figure}

\begin{figure}[h]
  \centering
     \includegraphics[width=3.4in]{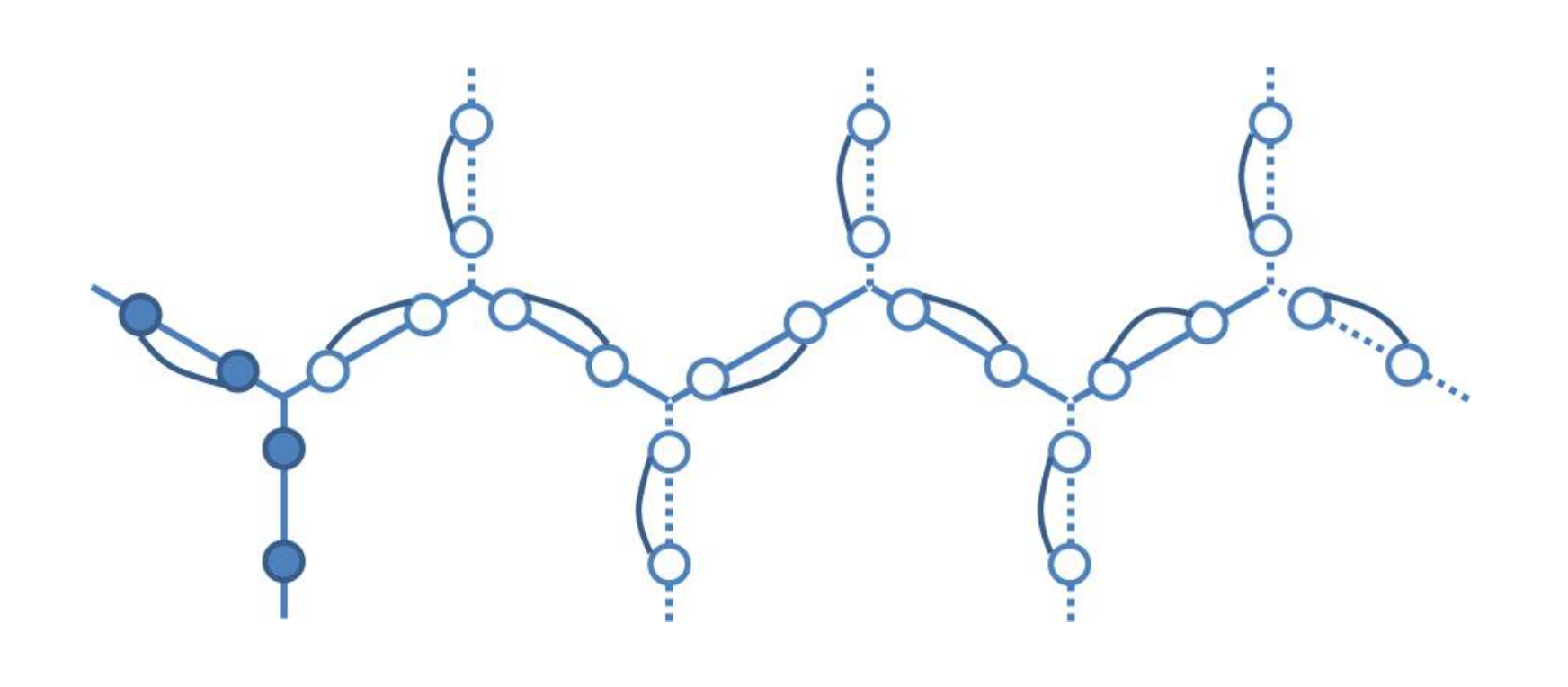}
	\caption{Possible configuration where no isolated spin $1/2$'s appear near the end of strings, showing that such excitations are confined.  The violated vertices are indicated with green arrows; red arrows indicate edges that are not in a ground state of $H_0$.  }
	\label{conf}
\end{figure}

Let us now discuss the spectrum of this new Hamiltonian.  By construction, any configuration in which all loops are closed will have lowest energy if all edges are in one of the two states $\ket{\tilde{0}}, \ket{\tilde{1}}$.  This is possible when $\prod \tau^z =1$ about each vertex: in this case each hard-core boson can have an associated spin $1/2$ at the vertex without violating the constraint; the possibilities are shown in Fig.\ref{dec}(a).  At vertices where  $\prod \tau^z =-1$, it is not possible for all edges to be in one of the two states $\ket{\tilde{0}}, \ket{\tilde{1}}$ without violating the constraint, as shown in Fig.\ref{dec}(b).  Hence whenever a string ends, the constraint that the total number of spins at each vertex must be even ensures that an unpaired spin $1/2$ remains at or near the vertex.  Moving this unpaired spin away from the vertex in question produces a series of edges along which $H_0$ is not minimized (Fig. \ref{conf}).  In other words, the unpaired spin is confined by a linear energy penalty to be close to the violated vertex.   Therefore an unpaired spin $1/2$ is bound (by a linear confining potential) to each end-point of an open string.  The corresponding semion string operator is obtained from Eq. (\ref{SString}) by substituting $\tau^x \rightarrow \tilde{\tau}^x$, {\it and} creating an extra spin-$1/2$ in the state $|\uparrow \rangle$ at each string end-point such that the total spin at each vertex remains integral.  

Defects in the original semion model involving only plaquette violations are unaffected by the decoration.  
 In particular, semions on the surface remain deconfined.  Because they are also bound by a confining potential to a spin-1/2, these deconfined surface semions transform projectively under the global symmetry,  exactly as in the Kalmeyer-Laughlin spin liquid.  In contrast excitations involving only plaquette violations in the undecorated Walker-Wang model transform linearly under the symmetry.\cite{Note3}

The construction given so far realizes a system with a Kalmeyer-Laughlin spin liquid state on its surface, by associating an open spin-singlet chain with the semion excitation.  
 Our main objective, however, is to obtain surface states that are not possible in 2 dimensions.  We can do this with a very minor modification of the construction given above: it suffices to take the spin $1/2$ degrees of freedom to transform under $\mathbb{Z}_2 \times \mathbb{Z}_2$ according to Eq. \ref{ps}.    In this case our spins are not really spins since we have assumed that $SO(3)$ spin rotation invariance is broken; however a spin singlet will still transform as a net singlet under symmetry.  Hence the operator $\tilde{\sigma}^x$ still interchanges two different trivial representations, and thus does not violate the symmetry.

\subsection{Global symmetries at the surface} \label{AKLTSurface}

Let us now verify that the surface of our 3D model behaves according to the expectations set out in Sect. \ref{anomaly}.  Specifically, we wish to understand how the fusion rules (\ref{VortFuse}) arise at the surface of our model.  

Our starting point will be the action of the symmetry $g$ (in the ungauged model) on a semion loop, which carries a singlet chain.  The singlet chain is essentially an AKLT chain in which we do not project onto spin $1$ at each site; like the AKLT chain, it has an exact matrix product ground state.   For a chain with $N$ sites and periodic boundary conditions, this has the form:
\be
|\Psi \rangle = \text{Tr} \left[  A^{(s^z_1)} ... A^{(s^z_N)} \right ] | s_1, ... s_N \rangle
\ee
where $s_i = 1, 0, -1$ for the three triplet states, or $b$ for the spin singlet state.  Here
\ba
A^{(1)} =& \frac{1}{\sqrt{2} } \sigma^+ \ , \ \ \  A^{(-1)} =& - \frac{1}{\sqrt{2} } \sigma^- \\
 A^{(0)} = & -  \frac{1}{2} \sigma^z \ , \ \ \  A^{(b)} =& -\frac{1}{2} \mathbf{1}
\ea
where $\sigma^{\pm} = \sigma^x \pm i \sigma^y$ with $ \sigma^{x,y,z}$ the Pauli matrices.  

The projective action of the symmetry can be unveiled using the approach of
Ref. \onlinecite{TurnerBerg}.  Consider the effect of the $\mathbb{Z}_2 \times \mathbb{Z}_2$ symmetry operations 
\be
R_{\alpha} = e^{ i \pi \hat{S}_\alpha} \ , \ \ \ \alpha = x,y,z
\ee
These act according to 
\ba
R_y | b\rangle & =&   | b \rangle \n
R_y | m_z = 1\rangle & =&   | m_z = - 1\rangle \n
 R_y | m_z = -1\rangle &=&   | m_z =  1\rangle \n
R_y | m_z = 0\rangle &= & -  | m_z = 0 \rangle 
\ea
and similarly for $R_x, R_z$.
In terms of our matrix product state, acting with $R_y$ on all sites in the chain must therefore take:
\ba
R_y: & \ A^{(1)} \rightarrow  A^{(-1)} \ , \ \ \ A^{(- 1)} \rightarrow  A^{(1)} \n  
& A^{(0)} \rightarrow - A^{(0)} \ , \ \ \ A^{(b)} \rightarrow  A^{(b)} 
\ea
These transformations are carried out by taking
\be
A^{(s)} \rightarrow U_y A^{(s)} U_y^\dag
\ee
with $U_y = \sigma^y$.  Similarly, $R_z$ acts on the matrix product state by conjugating all matrices $A^{(s_i)}$ by $U_z =\sigma^z$.  
For an open singlet chain, the first and last matrices are row and column vectors, respectively.  These transform via:
\be
A^{(s_1)} \rightarrow  A^{(s_1)} U_y^\dag \ , \ \ \ A^{(s_N)} \rightarrow U_y A^{(s_N)} 
\ee

Since the matrices $U, U^\dag$ appear in pairs, we are free to re-define $U_y$ and $U_z$ by arbitrary factors of $i$.  (Unlike in a 1D system, here we must require that both end-points of the chain transform in the same projective representation, so that $U_g^2$ must be real).   It is natural to take these phases such that $U_y^2 = U_z^2 = 1$.  As Ref. \onlinecite{TurnerBerg} emphasizes, however, once we have fixed the matrices $U_y, U_z$,   $U_x$ is fixed by the requirement that it be the product $U_y U_z$.  Specifically, since $(U_y U_z)^2 =-1$,    
\be
U_x = \pm i \sigma^x
\ee

\begin{widetext} 

\begin{figure}[htp]
  \centering
  \begin{tabular}{ll}
(a)&   (b)\\
  \includegraphics[width=3.0in]{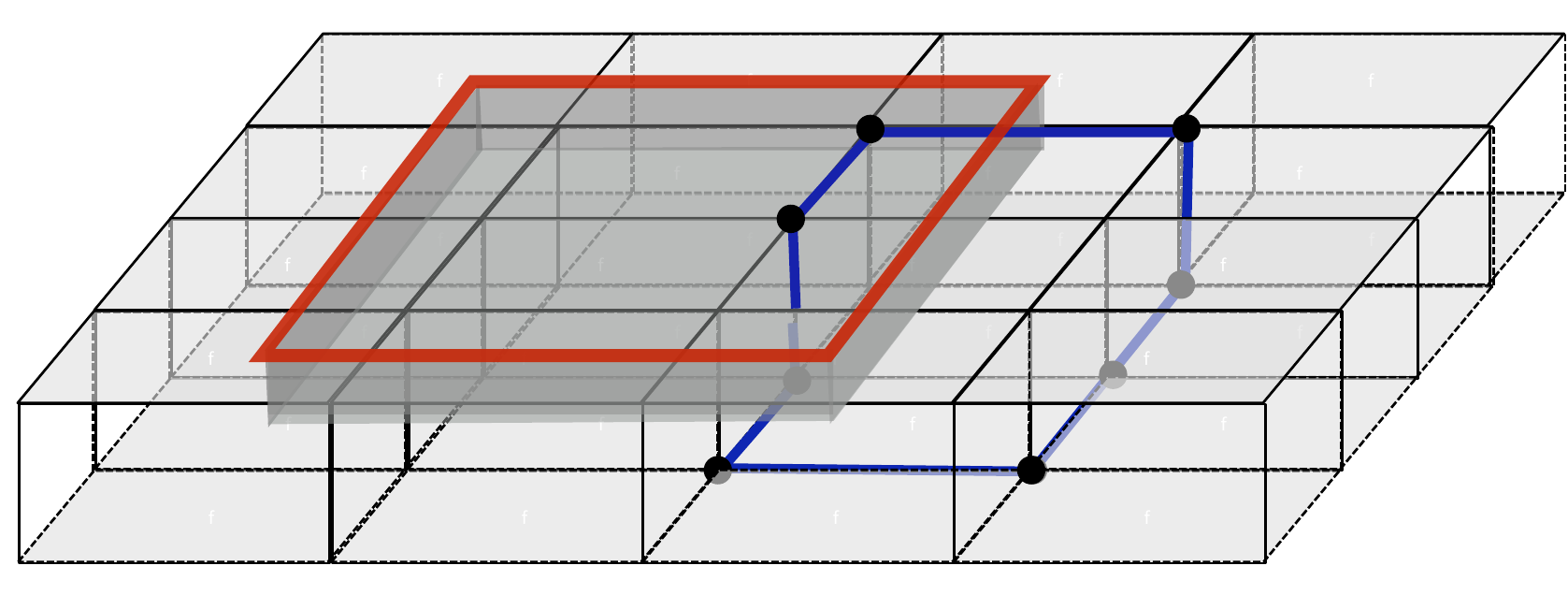} &   \includegraphics[width=3.0in]{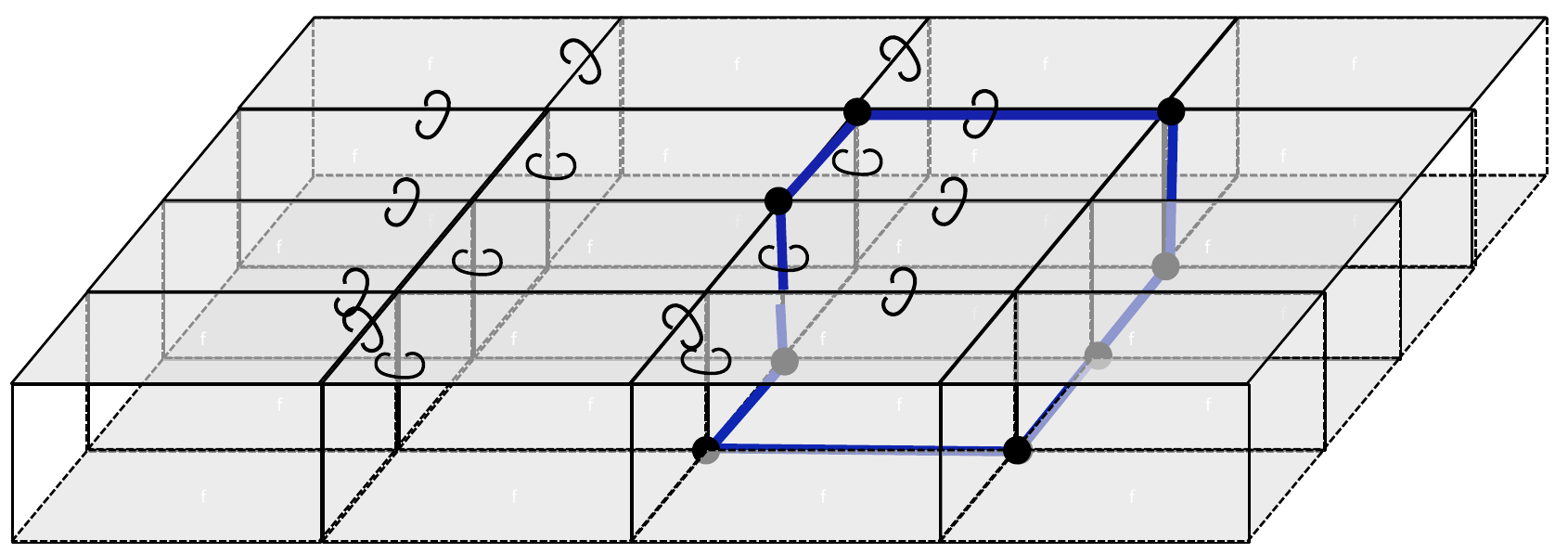} \\
  (c)&  (d) \\
     \includegraphics[width=3.0in]{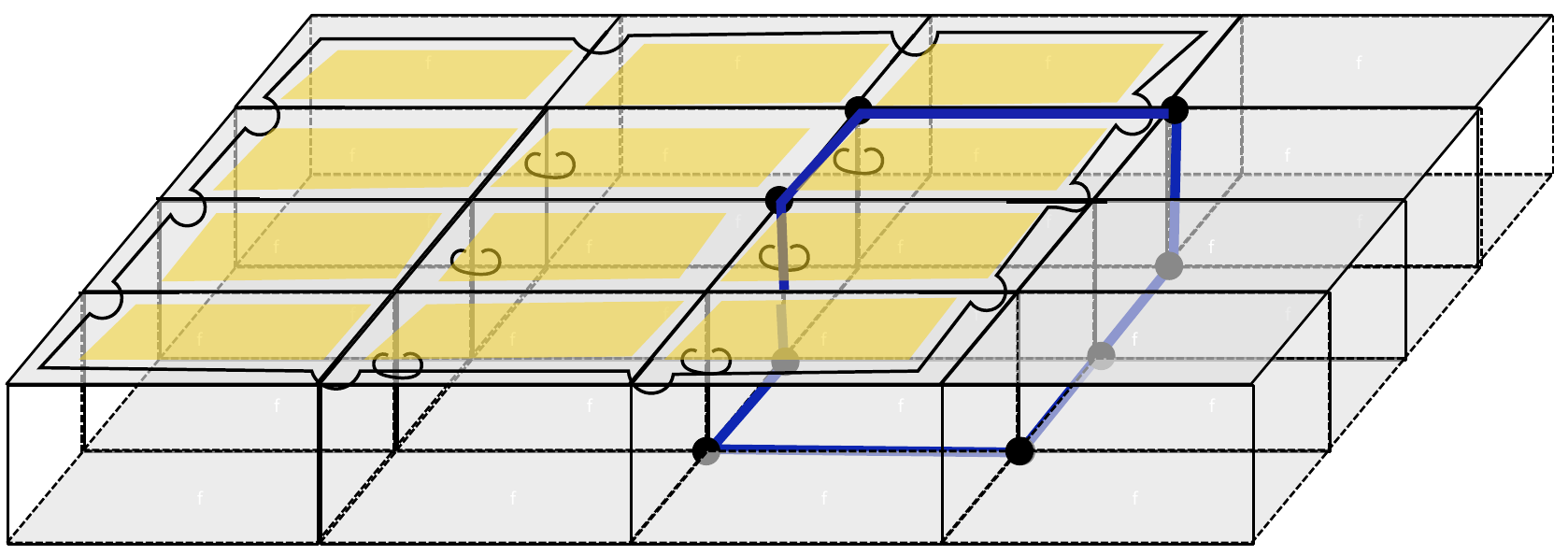} &   \includegraphics[width=3.0in]{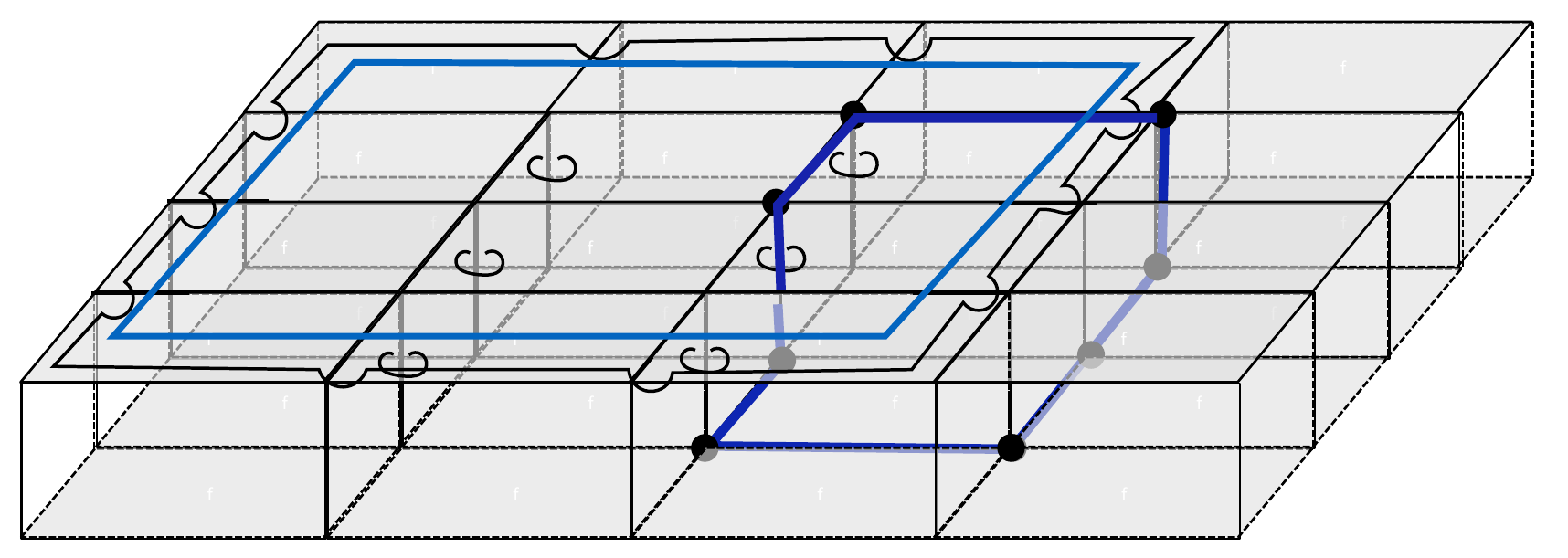} \\
       \end{tabular} 
	\caption{Symmetry action and surface semion strings.  (a) Acting with the symmetry operator $R_x$ on all spins inside the grey region creates a bulk domain surface (the boundaries of the grey box) that ends on a surface domain wall (shown in red).  This operation twists each singlet-chain link piercing the domain surface by $U_x$ (or $U^\dag_x$).  Performing this transformation twice multiplies the wave function by a phase factor $(-1)^{n_s}$, where $n_s$ is the number of edges piercing the domain surface that carry semion labels.   (b) Diagramatically, this can be represented by encircling each edge with a small semion loop.   (c) Acting with a product of plaquette operators on the shaded plaquettes reproduces this operator in the bulk, but leaves a large loop running below the surface along the domain wall.  (d)  To obtain the correct phase factor, an additional semion string (shown here in blue) must be added at the surface along the domain wall.  
Note that for simplicity, we have suppressed the point-splitting of the cubic lattice in this figure.	}
	\label{DWFig1}
\end{figure}

\end{widetext} 

To see how the states in our 3D model are affected by symmetry transformations, consider creating a domain wall on the 2D surface by applying the transformation $g$ at all sites inside the 3D region shown in Fig. \ref{DWFig1} a.  Suppose that an $N$ -site singlet chain enters and exits this region, such that domain walls are created between sites $(i-1,i)$ and between $(i+n-1, i+n)$.   After the symmetry transformation, the new matrix product state for the singlet chain is:
\be
\text{Tr} \left[  A_1 ... A_{i-1} U_g A_{i} ... A_{i+n-1} U_g^\dag  A_{i+n}...A_{N} \right ] | s_1, ... s_N \rangle  \nonumber 
\ee
where $A_1 \equiv A^{(s_1)}$, etc.   Hence the first domain wall is associated with an insertion of $U$, and the second with an insertion of $U^\dag$.  

Similarly, if a singlet chain ends inside the region, after the transformation its wave function is :
\be \label{PsiUd}
\text{Tr} \left[ A_1 ... A_i   U^\dag_g A_{i+1} A_{i+2} ...  A_N \right ] | s_1, ... s_N \rangle \n
\ee
where the domain wall occurs between sites $i$ and $i+1$.  Importantly, in this case $U_g^\dag$ appears only on one link in the chain.

If we act with $g=y,z$ twice on the same region, the resulting operator is simply the identity.  However, if we act twice with $g=x$, the result is
\be
\left( R_x(\mac{R}) \right )^2 = \prod_{i \in ^*\mac{R} } (-1)^{n_i} \equiv \Phi_s(\mac{R})
\ee
Here $\mac{R}$ is the 3D region in which the symmetry acts, and $^*\mac{R}$ is the set of edges that connect sites inside $\mac{R}$ to those outside (i.e. the set of edges bisected by the surface bounding $\mac{R}$).

The operator $\Phi_s$ can be expressed in a more illuminating form: on any surface $S$\cite{vonkeyserlingk13}
\be
\prod_S B_P = - W_{\partial S} \prod_{e \in ^*S } (-1)^{n_s(e)}
\ee
where $^*S$ is the set of edges that stick out of the surface $S$, but are not on the boundary of the 3D system.  $\partial S$ is the curve where $S$ intersects this boundary, and $W_{\partial S}$ is an operator creating a closed loop along this curve (Fig. \ref{DWFig1} c).\cite{Note4}

To turn this into the operator we want, we multiply it by a surface semion string operator $\hat{S}$ that runs along $\partial S$ (Fig. \ref{DWFig1} d):
\be \label{DegId}
\Phi_s = - \hat{S}_{\partial S} \prod_{P \in S} B_P 
\ee
where $S$ is the 2D surface bounding $\mac{R}$.\cite{Note5}

If we restrict our attention to situations where there are no defects in the 3D bulk of our model, then the product over plaquette operators acts as the identity, and acting with $\Phi_s$ is equivalent to acting with the surface semion string operator.  In this way, we see that acting twice with $R_x$ in $\mac{R}$ is equivalent to acting with a semion string that encircles the intersection of the boundary of region $\mac{R}$ with the surface of the 3D system.  Since acting with the symmetry is equivalent to braiding with an appropriate defect, this implies that two $\Omega_x$ defects will fuse to a semion string.  Similar arguments can be used to deduce the appearance of the other coefficients in the fusion relations (\ref{VortFuse}).

\begin{figure}[ht]
  \centering
  \begin{tabular}{l}
 (a)\\    \includegraphics[width=3.2in]{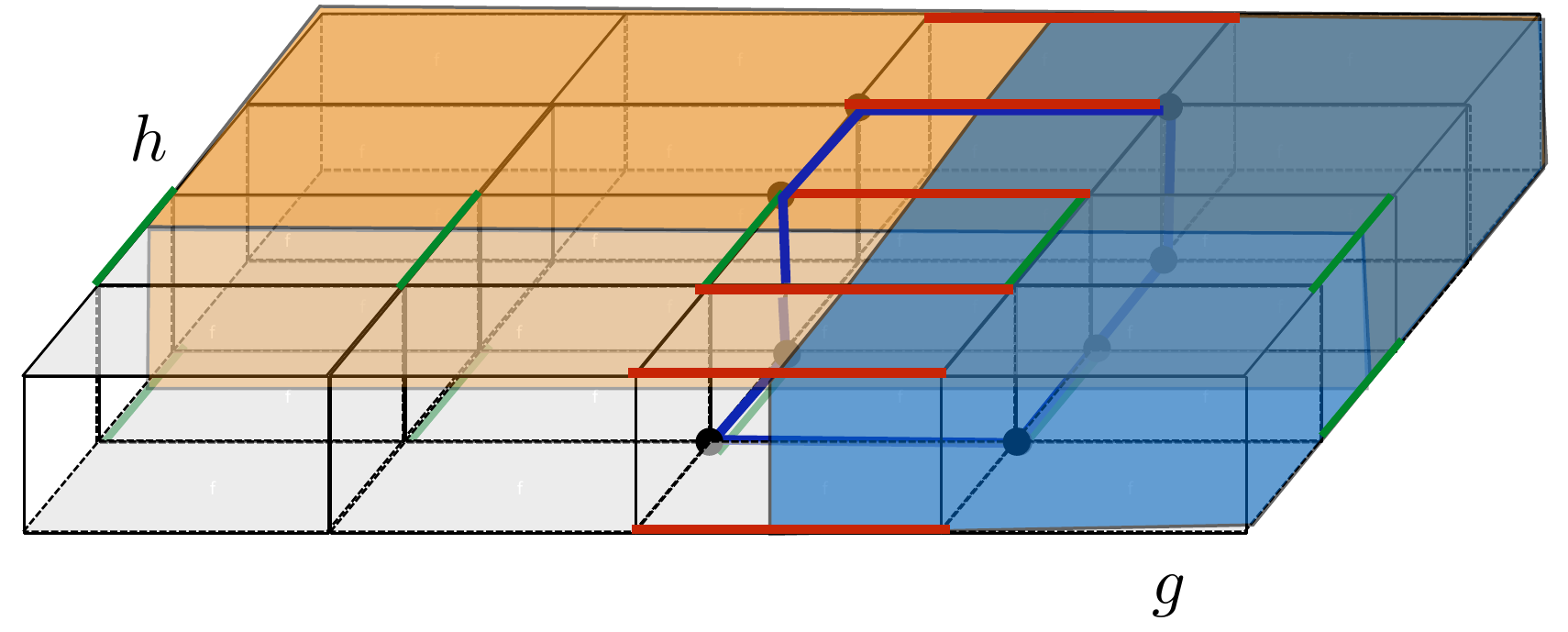} \\
(b)\\       \includegraphics[width=1.6in]{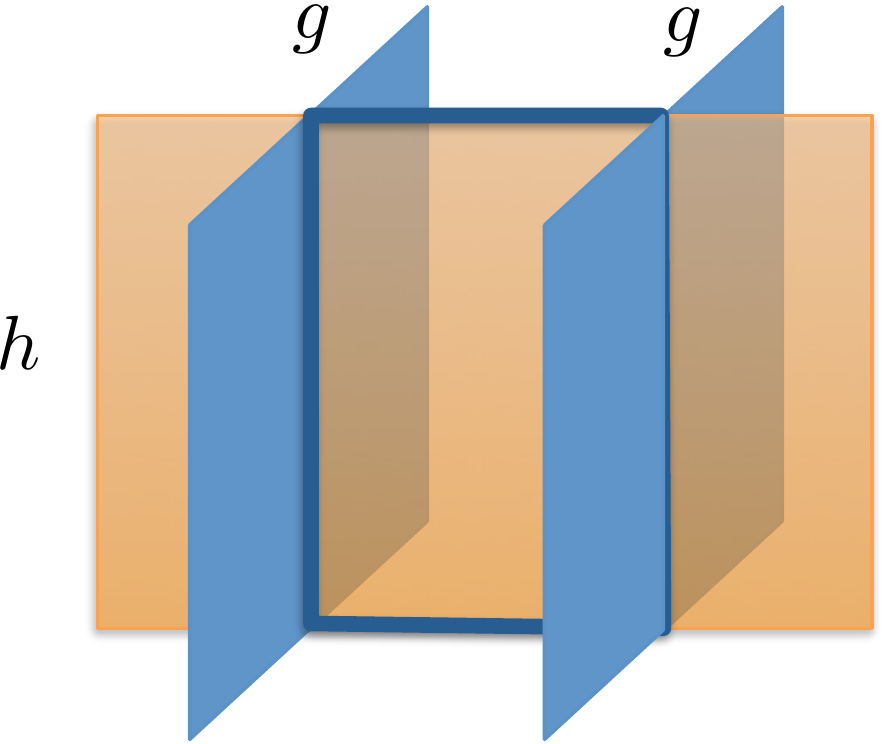} 
\end{tabular}
	\caption{Intersecting pairs of domain walls: (a) Here we first act with $g$ on all spins inside the blue shaded region, twisting all singlet chains along the set of red edges by $g$.  We next act with $h$ on all spins inside the mauve shaded region.  This will twist the singlet chains on a second set of bonds (shown in green) by $h$, and also generate a phase factor of $(-1)^{n_s}$ for each red edge inside the mauve region.  (b) When acting on states where $B_P \equiv 1$, this is equivalent to running a semion string (thick blue line) around the boundary. }
	\label{DWFig2}
\end{figure}

\subsection{Intersecting domain walls} \label{GdwC}

Next, consider the effect of creating first a pair of $g$ domain walls, and then a pair of $h$ domain walls in an orthogonal plane (Fig. \ref{DWFig2} a).  After making the pair of $g$ domain walls, the edges bisecting each domain wall have been ``$g$-twisted": if a semion loop crosses the $g$ domain wall on edges $(i-i, i)$ and $(j, j+1)$, the spin component of the wave function on this loop is now
\be \label{}
\text{Tr} \left[  A_1 ... A_{i-1} U_g A_{i}   ...A_{j} U_g^\dag A_{j+1} ...  A_N  \right ] | s_1, ... s_N \rangle 
\ee

Next, we create a pair of $h$ domain walls.  
This does two things.  First, it twists a second set of edges by $U_h$.  Second, it acts on the $g$- twisted edges between the $h$ domain walls to send
\be
 U_g \rightarrow U_h U_g U^\dag_h 
 \ee
For $h \neq g$, this gives an extra phase factor of $-1$ for each $g$-twisted edge in the region acted on by $h$.  This reflects the fact that these edges are no longer singlets and therefore carry (integer) charge under the remaining generators of $\zt \times \zt$. 

 If $g \neq h$, we thus obtain an overall phase factor $\prod (-1)^{n_s}$, where the product runs over all links twisted by $g$ and then acted on by $h$.  Exactly as in the case of the domain surface, this phase can be represented by a product of plaquette terms, combined with a semion string running along the boundary of the intersection between the domain walls (Fig. \ref{DWFig2} b).  

 Hence in our exactly solvable model, the action of symmetry is intertwined with the topological degrees of freedom in two ways.  First, appropriate combinations of global symmetries in 3D regions that end on the surface are (provided the bulk is in its ground state) equivalent to semion string operators, as anticipated from the discussion of section \ref{anomaly}.  Second, such semion strings also appear at intersections of different types of domain walls.  In the next section, we will use the second result - re-derived there in a slightly different way in the context of the gauged theory - to show that gauging the symmetry results in 3-loop braiding statistics.

\section{Non-trivial nature of our 3D SPT via 3 loop braiding}
\label{3loopbraiding}

In the previous section, we constructed a 3d Hamiltonian invariant under the symmetry $G = \zt \times \zt$ with a trivial bulk and a surface state that is anomalous, by the criterion of Sec. \ref{anomaly}.  We can thus conclude that this must be a non-trivial 3d SPT, for otherwise we could decouple the bulk and have the anomalous surface state exist as a standalone 2d theory, an impossibility.  However, it is nice to have a direct verification of the non-trivial SPT nature of our 3d bulk, especially since robust diagnostics of nontrivial 3d SPT order have been developed \cite{Wang2014a, Jiang2014}.  In this section we will show that (a slight modification of) our 3d model is indeed a non-trivial 3d SPT according to these criteria.  Specifically, we gauge  $G$ in the 3d SPT model, and show that flux insertion operators in the gauged theory satisfy a non-trivial algebra \cite{Jiang2014}.  The non-trivial algebraic relation we show can also be interpreted in terms of 3-loop braiding \cite{Wang2014a}.

In order to gauge the theory, it is convenient to introduce a more abstract (but more general) variant of the decorated WW model discussed above.  It can easily be checked that when $G$ is global, this alternative model reproduces the phenomena discussed in sections \ref{AKLTSurface} and \ref{GdwC}: namely, semion lines appear at the boundaries of domain surfaces, and encircle the intersections of pairs of domain walls of different type.  Because of this, we only give a schematic description of the features we will require.

Our new Hamiltonian is again that of a generalized spin system, with generalized spins at each vertex of the cubic lattice.  The spin Hilbert space at a vertex $V$ is
\be
\calH_V=\calH_V^G \otimes \calH_V^{WW}
\ee
where $\calH_V^{WW}$ is the Hilbert space of all possible semion labelings of the edges adjacent to $V$ that satisfy the vertex term $A_V=1$, and $\calH_V^G$ is a tensor product of `parton' Hilbert spaces ${\mathbb{C}}^{|G|} = {\mathbb{C}}^4$ over these edges ($G=\mz_2 \times \mz_2$).  Note that in a general state, adjacent vertex Hilbert spaces may have disagreeing labels for the edge they share; the first term we impose in the Hamiltonian is an energetic penalty for this scenario, with the rest of the terms in the Hamiltonian acting as $0$ unless this label matching condition is satisfied.  From now on we implicitly work in this constrained Hilbert space - but note that surface semion quasiparticle excitations are violations of this constraint.

Each site Hilbert space $\calH_V$ is a linear unitary representation of $G$, with $f\in G$ acting by $U_V(f)$, defined by

\be
U_V(f)|V;{\bf g},{\bf s}\rangle = \chi(f, {\bf g}, {\bf s}) |V;f{\bf g},{\bf s}\rangle,
\ee
where the phase factor is defined in terms of our 2-cocycle $\omega$
\be
 \chi(f, {\bf g}, {\bf s})=\prod_{V' \tilde V} \omega({\bf g}(V')f,f)^{{\bf s}(V')}.
 \ee
Here ${\bf g}(V') \in G$ represents the parton degree of freedom at site $V$ corresponding to the edge $\langle V V'\rangle$, and ${\bf s}(V')=0,1$ is the semion occupation of link $\langle V V' \rangle$.  This can be seen to be a linear representation thanks to $A_V=1$.

We now define the Hamiltonian 
\be
H=H_{constr.}+ H_G+H_{WW}
\ee
as a sum of local terms as follows.  $H_{constr.}$ contains the  constraint that semion labels for vertices $V$ and $V'$ at either end of the edge $\langle V V'\rangle$ should agree.  The rest of the terms are defined to act as $0$ unless the labeling constraints are satisfied in their vicinity.  If they are satisfied, $H_G$ is defined as follows.  It acts purely on the parton degrees of freedom, and is a decoupled sum over edges $\langle V V' \rangle$ of projectors onto the state $\frac{1}{\sqrt{|G|}} \sum_g |g\rangle \otimes |g\rangle$, where the two tensor factors correspond to the $G$ partons associated to $V$ and $V'$ respectively.  We can again think of these projectors as constraints.  Furthermore, $H_{WW}$ is precisely the un-decorated Walker-Wang Hamiltonian discussed above, with the caveat that the vertex and plaquette terms in it act as $0$ unless both the labeling and parton constraints are satisfied in their vicinity.  This definition makes it easy to check that all terms commute, and a ground state satisfying all constraints exists.  Indeed, this ground state is clearly just the ground state of the original Walker-Wang model, tensored with the state where the $G$ partons are entangled as $\frac{1}{\sqrt{|G|}} \sum_g |g\rangle \otimes |g\rangle$ along all the links.  Note that the only coupling between $H_G$ and $H_{WW}$ is through the definition of the action of $G$, and specifically the phase $\chi$ defined above.

\subsection{Membrane algebra: generalities}

To extract the membrane algebra invariant of reference \onlinecite{Jiang2014}, we couple our model to a $G$ gauge field.  We do this by introducing $G$-valued variables living on links (note that we do not need an orientation because everything is $\mz_2$ valued) and minimal coupling.  The precise meaning of minimal coupling is as follows.  First of all, we give an energetic penalty for non-trivial $G$ flux through any plaquette.  Second, we define $G$-coupled versions of the local terms in $H_G$ and $H_{WW}$ by having them act as $0$ unless there is no $G$-flux through any of the plaquettes in the vicinity of the local term in question, and in the case when there is no such $G$-flux, we define them by having them act in the unique way dictated by $G$ gauge invariance.  Finally, we add terms for the $G$ gauge field variables which perform $G$ gauge transformations on vertices - {\emph{i.e.}} vertex terms for the $G$ electric flux.  The entire model stays exactly solvable and is in the deconfined phase of the $G$ gauge theory.  Our goal is to study the algebra of membrane flux insertion operators in it.

For now, we will write everything in terms of the two-cocycle $\omega$ defining our model; later we will specialize to a particular $\omega$.  Following reference \onlinecite{Jiang2014}, our geometry will be a three-torus whose dimensions we refer to as $x,y,z$ (note the lower case letters).  Let us first construct a membrane flux insertion operator ${\bf Z}_{xy}$ that inserts $Z$ flux in the $xy$ plane.  Recall our notation $\{I,X,Y,Z \}$ for the group $G= \mz_2 \times \mz_2$.  The requirements for ${\bf Z}_{xy}$ are that it act only on the degrees of freedom (in the gauged model) in the vicinity of the $xy$ plane, that it map the $|G|^3=64$-fold degenerate ground state space of the model on the three torus to itself while inserting flux $Z$ in the $z$ direction, and that $({\bf Z}_{xy})^2$ act as the identity on this ground state space.  Note that these conditions do not uniquely determine the action of ${\bf Z}_{xy}$ in the ground state space: we can always multiply any valid ${\bf Z}_{xy}$ by operators that measure $G$-flux along the $x$ and $y$ directions.  The criterion of reference \onlinecite{Jiang2014}, which is constructed in terms of such membrane flux insertion operators and determines whether we have a non-trivial SPT or not, is of course insensitive to such ambiguities.

\begin{figure}[h]
  \centering
     \includegraphics[width=3in]{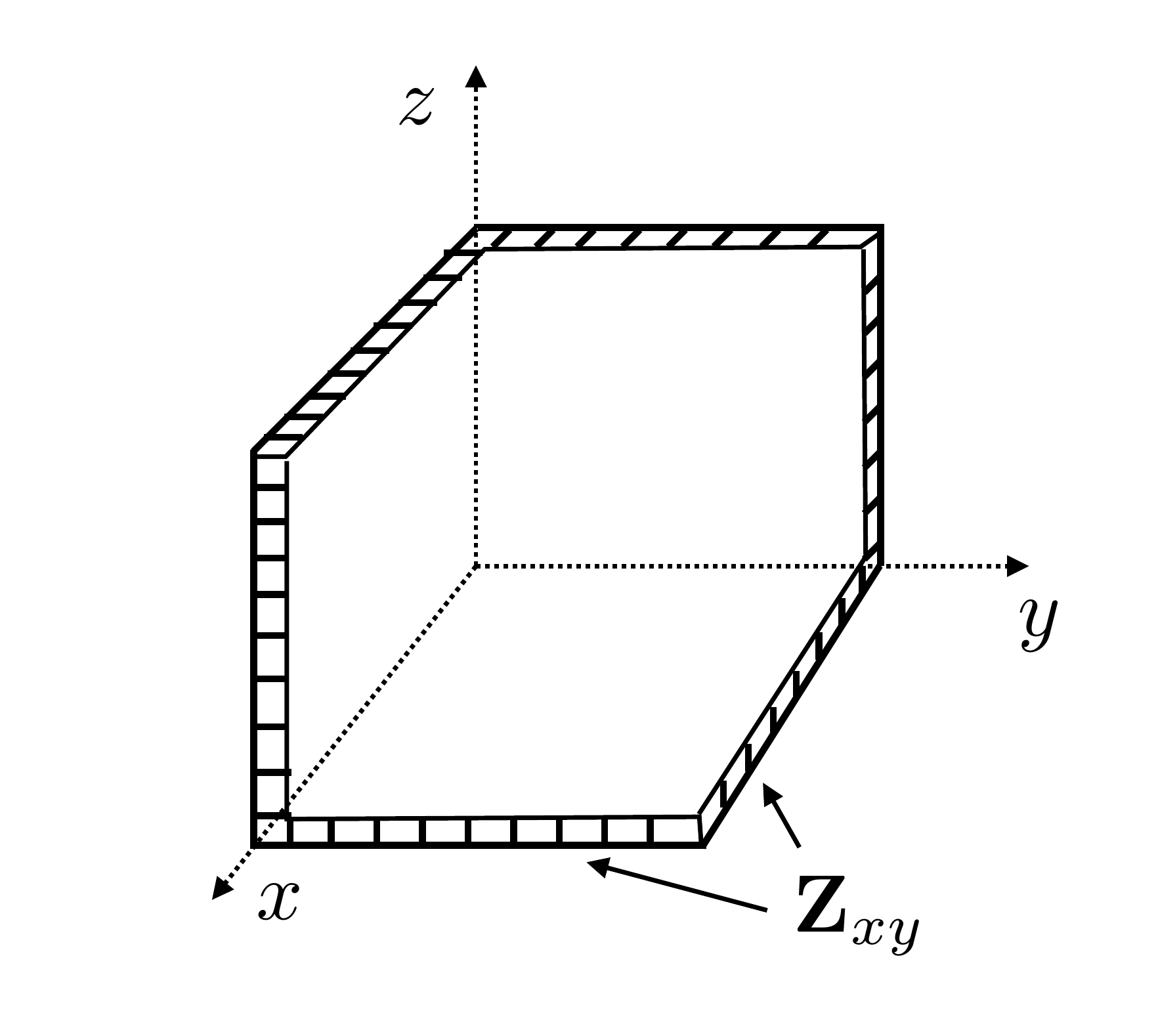}
	\caption{The operator ${\bf Z}_{xy}$ acts on the gauge field by changing by $Z \in G$ the values of the gauge field variables on the indicated links; in other words, it inserts a $Z$ flux in the $z$ direction.  Similarly, operators that insert flux in the other directions act in the other planes.}
	\label{LF1}
\end{figure}

We now construct a particular ${\bf Z}_{xy}$:

\be
{\bf Z}_{xy} = Z^{gf}_{xy} {\hat Z}_{xy}
\ee
where $Z^{gf}_{xy}$ will act only on the gauge field degrees of freedom and ${\hat Z}_{xy}$ will act on the original spin degrees of freedom and depend on but not change the gauge field degrees of freedom.  $Z^{gf}_{xy}$ is defined by taking a given gauge field configuration and changing by $Z$ the gauge field variables living on the vertical links $\langle V V'\rangle$, where $V$ is a link of the $xy$ plane and $V'=V+{\hat z}$ - see figure \ref{LF1}.  As for ${\hat Z}_{xy}$, we define it as follows.  For each of the $4^2=16$ $x$ and $y$ direction $G$ flux sectors, we pick a particular $G$ gauge field configuration, and define ${\hat Z}_{xy}$ in that sector by a particular operator written out below.  We then extend it uniquely to an operator in the full gauge theory using the requirement of gauge invariance.  Note that we can define ${\hat Z}_{xy}$ arbitrarily in each $x$ and $y$ (but not $z$) flux sector, since ${\hat Z}_{xy}$ can measure the $G$ holonomy in the $x$ and $y$ direction.  The requirement of gauge invariance is sufficient to guarantee that ${\bf Z}_{xy}$ will map ground state to ground state in the gauged theory, provided that ${\hat Z}_{xy}$ does so for the un-gauged model.  

Now, the criterion of reference \onlinecite{Jiang2014} states that the zero flux ground state expectation value of

\be
{\bf Z}_{xy}^{-1} {\bf Z}_{xz}^{-1} {\bf X}_{yz}^{-1} {\bf Z}_{xz} {\bf X}_{yz} {\bf Z}_{xy}
\ee
is equal to $\pm i$ in the anomalous theories, and $\pm 1$ in the non-anomalous theory.  To compute this for our models, we look for the phase difference between
\be
\label{order1}
{\bf Z}_{xz} {\bf X}_{yz} {\bf Z}_{xy}|\psi_0 \rangle
\ee
and
\be
\label{order2}
{\bf X}_{yz} {\bf Z}_{xz} {\bf Z}_{xy}|\psi_0 \rangle,
\ee
where $|\psi_0\rangle$ is the zero flux ground state.  Note that in this computation we can ignore many signs, since we can tolerate sign errors in looking for a phase difference of $\pm i$.  By the discussion above, this amounts to evaluating the corresponding products of operators ${\hat Z}_{xy}$, etc. in the un-gauged theory, while separately keeping track of the gauge field dictated by the definition of our bold faced operators {\it in a particular gauge}.  A convenient gauge choice is to allow the gauge field configurations to be nonzero only on the $\langle V V'\rangle$ links where $V$ is in one of the cardinal planes and $V'$ is in a fixed normal direction away from it - see figure \ref{LF1}.  It is for these specific gauge field configurations that we will define the ${\hat X}, {\hat Z}$ operators, as described above.  Since the gauge field configurations end up the same regardless of whether we pick order \ref{order1} or order \ref{order2}, we just need to focus on the ${\hat{X}},{\hat{Y}},{\hat{Z}}$ operators in figuring out the phase difference.

\subsection{Membrane algebra for the anomalous and non-anomalous projective semion models}

We will now focus on the $Y$ anomalous projective semion model, defined by the $2$-cocycle $\omega_Y$, as well as the non-anomalous projective semion theory with $2$-cocycle $\omega_I$.  For completeness, recall that both of these $2$-cocycles are valued in $\pm 1$, with $\omega_Y=-1$ on $(Y,Y),(X,Y),(Y,Z),(X,Z)$, and $\omega_I=-1$ on $(X,X),(Y,Y),(Z,Z),(X,Y),(Y,Z),(Z,X)$.

According to our discussion above, we have to define 5 distinct operators in the un-gauged theory: ${\hat Z}_{xy}$ in the zero flux sector, ${\hat X}_{yz}$ and ${\hat Z}_{xz}$ in the sector with $Z$ flux in the $z$ direction and zero flux elsewhere, ${\hat Z}_{xz}$ in the sector with $Z$ flux in ${\hat z}$ and $X$ flux in ${\hat x}$, and ${\hat X}_{yz}$ in the sector with $Z$ flux in ${\hat z}$ and $Z$ flux in ${\hat y}$.  Each of these  will split into a product of two operators: one that just acts on the $G$ degrees of freedom, and one that acts on the Walker Wang degrees of freedom while depending on the $G$ degrees of freedom: ${\hat A} = {\hat A}_{G} {\hat A}_{WW}$.  Let us first describe the case of ${\hat Z}_{xy}$, where only ${\hat Z}_{xy,G}$ is non-trivial, and then see why in the other cases, a portion that acts non-trivially on the WW degrees of freedom is necessary.

We define ${\hat Z}_{xy}={\hat Z}_{xy,G}$ by having it introduce twists along the links $\langle V V' \rangle$, where $V$ is in the $xy$ plane and $V'=V+{\hat z}$.  Specifically, it acts only on the parton Hilbert space ${\mathbb C}^{|G|}$ associated to vertex $V$ in the $\langle V V'\rangle$ direction by

\be
|g\rangle \rightarrow \omega(gZ,Z) |g\rangle
\ee
where $\omega=\omega_Y$ in the anomalous projective semion $Y$ theory, and $\omega=\omega_I$ in the non-anomalous CSL.  It is straightforward to check that ${\hat Z}_{xy,G} |\psi_0\rangle$ is the ground state of the Hamiltonian with a $Z$ twist in the $xy$ plane: one just needs to check that it satisfies the twisted plaquette and $H_G$ terms.

The remaining operators will have a non-trivial Walker-Wang part, as one might expect from the discussion of intersecting domain wall pairs in Sect. \ref{GdwC}.  The reason is much the same as in the global case: a semion string passing through a $Z$ flux carries $X$ charge, so that applying $X$ globally in between two offset $yz$ planes results in an extra factor of $(-1)$ raised to the power of the number of semion strings passing the $Z$ plane between these two offset planes (Fig. \ref{DWFig2}).   This sign can be accounted for, using the linking sign property of semion strings, by running extra semion strings along the $y$ axis direction at the intersections of the offset planes with the $xy$ plane.

\begin{figure}[h]
  \centering
     \includegraphics[width=2.5in]{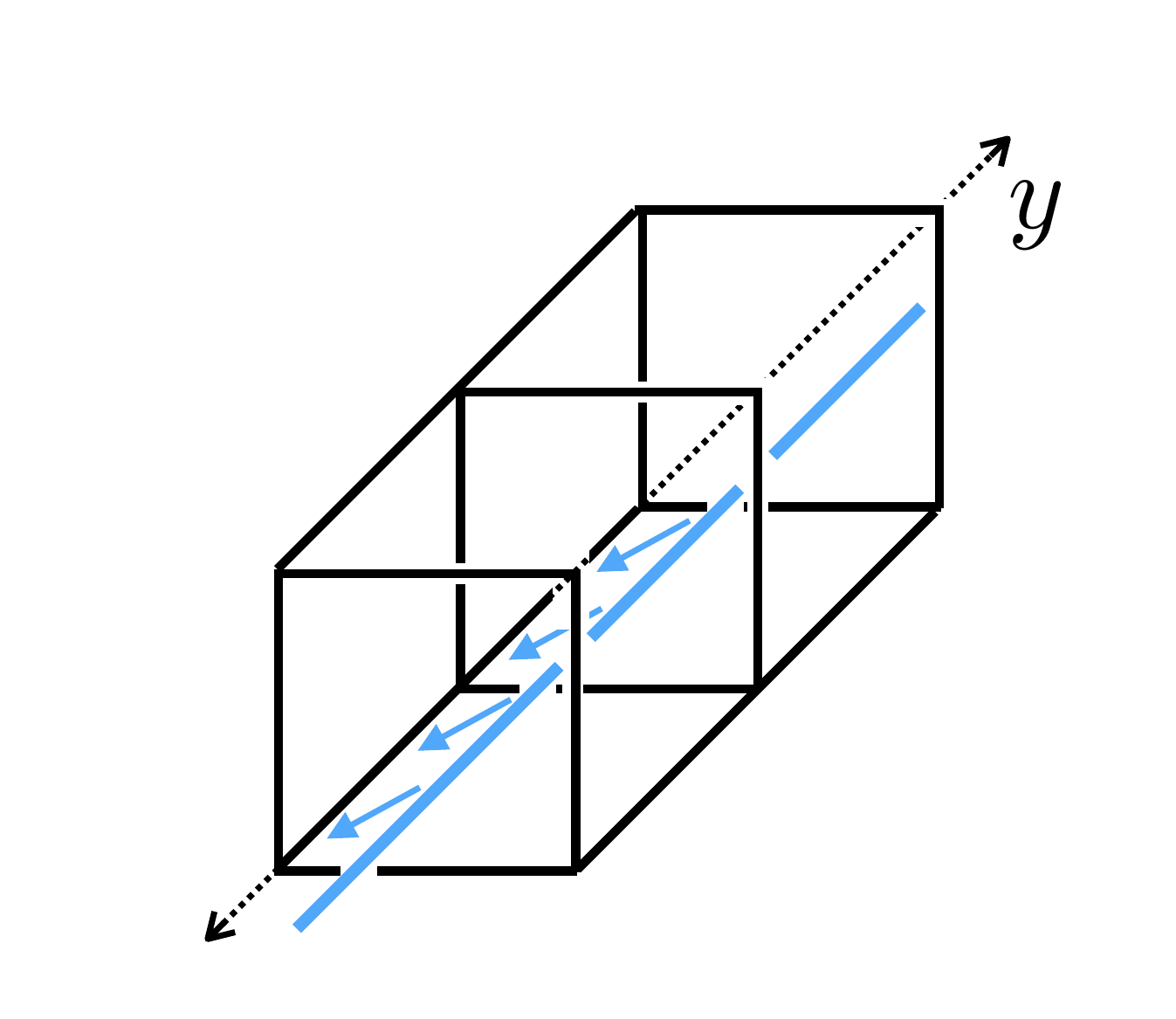}
	\caption{The Walker-Wang part of the membrane operator that inserts a flux of $X$ on a state which already has a flux of $Z$ in a perpendicular direction.  Naive successive application of the twist operators defined in the text leave a line of violated plaquettes at the intersection of the two fluxes.  These are satisfied by running a semion string operator (blue line) through them.  This blue semion string is glued into the lattice skeleton of our model using the local Walker-Wang graphical rules.}
	\label{LF2}
\end{figure}

 In the gauged model, when we act say with an $Z$ flux insertion operator in the $xy$ plane, followed by an $X$ flux insertion operator in the $yz$ plane, applying the above twist operators in succession produces a state which satisfies all terms in the Hamiltonian \emph{except} a line of $xz$ plaquette terms along the $y$ axis. 
To get an operator that maps ground state to ground state we need to get rid of these violated plaquettes, and we do so by including an extra semion string operator threading the line of violated plaquettes, illustrated in figure \ref{LF2}.  This is the extra Walker-Wang part of the flux insertion operators.  The existence of this line of violated plaquettes is of course another manifestation of the result of the previous section.

We can think of such a string insertion operator heuristically as follows: we take an extra semion string threading the line of violated plaquettes, and fuse it into the $y$ axis using the picture calculus rules that define the Walker-Wang model.  This is simply the product of operators $\tau^x$ along the links of the $y$ axis ($\tau^x$ is the Pauli operator that changes the semion occupation number), dressed by a sign depending on the occupation of the nearby links.  To satisfy the $H_G$ part of the Hamiltonian, there is also a sign that depends on the parton variables.  

We will actually not need an explicit expression for the semion string insertion operator, but only one key property: namely, two such string insertion operators along perpendicular directions must anti-commute.  This just follows from the pictorial Walker-Wang rules: the order of operations can be switched by sliding one semion line past the other, incurring a $-1$ from the linking.  It can also be seen from an explicit expression for the operators.

\subsection{Explicit computation}

\begin{figure}[h]
  \centering
     \includegraphics[width=3.2in]{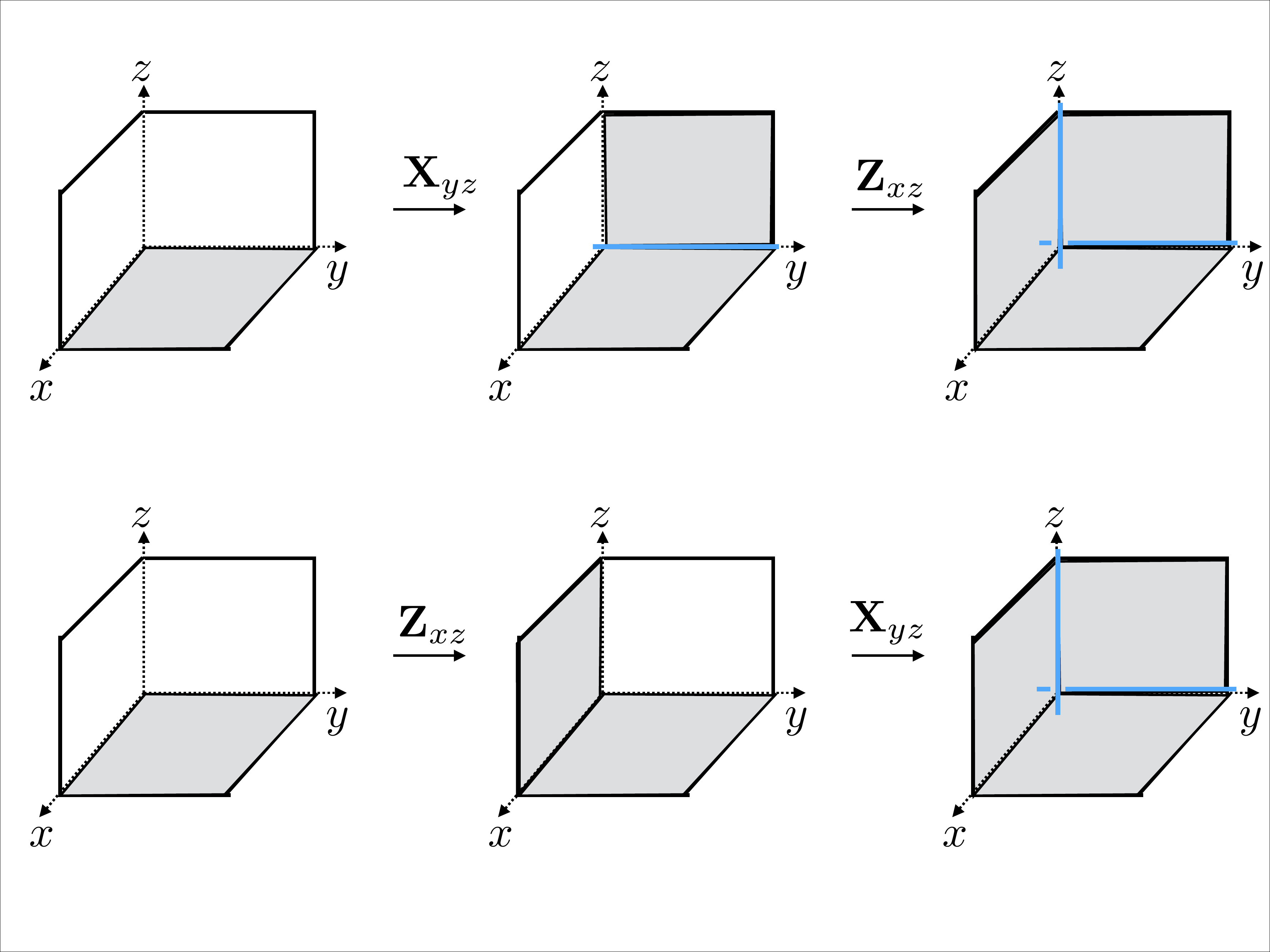}
	\caption{Graphical illustration of the two orderings, eqs. \ref{order1} and \ref{order2}.  They result in two states that differ by $\pm 1$ in the non-anomalous model, but $\pm i$ in the anomalous projective semion model $Y$ theory, as explained in the text.}
	\label{LF3}
\end{figure}

Armed with these flux insertion operators, we can now analyze the two models.  The form of the flux insertion operators is actually the same in the two cases of anomalous Y projective semion model and the non-anomalous CSL model.  The only difference is in the co-cycles, which enter the twist, and overall phase factors.  These overall phase factors turn out to be crucial, and can be obtained, up to sign, from the condition that all the flux insertion operator square to $+1$.

In figure \ref{LF3} we illustrate the two orderings of flux insertion operators.  Let us focus on the operator ${\bf X}_{yz}$ in the lower sequence in the figure.  It inserts two semion strings, one along the $y$ axis and one along the $z$ axis, in that order.  We claim that its coefficient must be $\pm i$.  To see this, note that without it, the operator would square to $-1$.  This $-1$ comes entirely from the Walker-Wang part of the operator, which involves inserting a semion string along $y$, then along $z$, then along $y$, and then along $z$ again.  These strings all cancel, but in doing so, acquire a $-1$ when a string along $z$ moves past the string along $y$.  Hence we need a factor of $i$ in front of this operator.

In the anomalous projective semion model, this factor of $i$ is the whole story: doing the flux insertion the other way (${\hat X}_{yz}$ first, then ${\hat Z}_{xz}$) results in a wave function whose amplitudes are all real (assuming a choice of overall phase that makes $|\psi_0\rangle$ have all real amplitudes), so the two orderings yield wave functions - which must be the same up to phase due to uniqueness of the SPT ground state - that differ by a factor of $\pm i$.

In the non-anomalous projective semion theory, there is however another factor of $i$, in addition to the one discussed above, so the ultimate phase difference between the two states is $\pm 1$, as expected.  This is the overall phase factor in front of ${\hat Z}_{xz}$ acting on the state with $Z_{xy}$ and $X_{yz}$ fluxes in place - that is, the operator corresponding to ${\bf Z}_{xz}$ in the upper sequence of figure \ref{LF3}.  While in the anomalous $Y$ model this phase factor is simply $\pm 1$, in the non-anomalous model it must be $\pm i$.  This is because without it, the square of ${\hat Z}_{xz}$ would be $-1$, owing to the semion string running along the $y$ axis of the state on which it is acting (middle frame of upper sequence in figure \ref{LF3}).  More precisely, it is because there are an odd number of such semion strings intersecting the $xz$ plane - the semion string is of course not pinned in any way to the $y$ axis.  The square of ${\hat Z}_{xz}$ on a  state with an odd number of semion lines is $-1$ due to the form of the co-cycle $\omega_I$ - it is just the statement that $Z^2=-1$ at the endpoint of a semion string in the non-anomalous model.  Thus we need a factor of $\pm i$ in front of this operator to have something that squares to $+1$, and so finally the phase difference between the two expressions eq. \ref{order1} and eq. \ref{order2} is just a sign, as expected in a non-anomalous theory.

\section{$O(5)$ Non-linear sigma model construction of our SPT and anomalous surface}  
\label{O5}

An alternative approach to constructing SPTs is to use non-linear
sigma models (NLSMs) with a theta term. In particular, Ref.\onlinecite{Bi2013,Vishwanath2013} argue that a wide class of 3d SPTs can be described by the following action:

\begin{center}
$S_{3d}=\int d^{3}x\, d\tau\,\frac{1}{g}(\partial_{\mu}\vec{n})^{2}+S_{\theta}$
\par\end{center}

\noindent where

\begin{center}
$S_{\theta}=\frac{2\pi i}{64\pi^{2}} \int d^{3}x\, d\tau\, \epsilon_{abcde}\epsilon^{\mu\nu\rho\delta}n^{a}\partial_{\mu}n^{b}\partial_{\nu}n^{c}\partial_{\rho}n^{d}\partial_{\delta}n^{e}$
\par\end{center}

\noindent Specifically, for $G=\mathbb{Z}_{2}^{A}\times\mathbb{Z}_{2}^{B}$,
Ref.\onlinecite{Bi2013} argues that the following action of $G$ realizes a non-trivial
$G$-SPT:

\begin{center}
$\mathbb{Z}_{2}^{A}:n_{1,2}\rightarrow-n_{1,2},n_{a}\rightarrow n_{a}(a=3,4,5)$
\par\end{center}

\begin{center}
$\mathbb{Z}_{2}^{B}:n_{1}\rightarrow n_{1},n_{a}\rightarrow-n_{a}(a=2,3,4,5)$
\par\end{center}

\noindent We will see how to construct one of our anomalous semion
surface states as a symmetric termination of this 3D SPT.

Before we do this, let us first consider an alternative action of
this symmetry group, corresponding to a trivial 3D SPT.
For clarity, we will denote this group by $\tilde{\mathbb{Z}}_{2}^{A}\times\tilde{\mathbb{Z}}_{2}^{B}$.
Its action is:

\begin{center}
$\tilde{\mathbb{Z}}_{2}^{A}:n_{1,2}\rightarrow-n_{1,2},n_{a}\rightarrow n_{a}(a=3,4,5)$
\par\end{center}

\begin{center}
$\tilde{\mathbb{Z}}_{2}^{B}:n_{2,3}\rightarrow-n_{2,3},n_{a}\rightarrow-n_{a}(a=1,4,5)$
\par\end{center}

\noindent Now, this is simply the subgroup of 180 degree rotations
inside the $SO(3)$ that rotates $n_{1},n_{2},n_{3}$, and since we
know that there is no non-trivial 3D SPT of $SO(3)$ in 3D, the corresponding
$\tilde{\mathbb{Z}}_{2}^{A}\times\tilde{\mathbb{Z}}_{2}^{B}$ SPT
must be trivial as well. But let us examine its surface anyway. To
do this, consider the effective action for the complex field $n_{4}+in_{5}$
on the 2D surface, after having integrated out $n_{1,2,3}$. According
to the arguments of Ref.\onlinecite{Bi2013,Vishwanath2013}, the theta term ensures
that a single vortex of $n_{4}+in_{5}$ carries a spin $1/2$. Now,
we can add fluctuations and proliferate doubled vortices bound to
single charges - i.e. we think of this as a system with $U(1)$ charge
conservation symmetry that acts on $n_{4}+in_{5}$, and drive it to
a $\nu=1/2$ bosonic Laughlin state, while maintaining the $SO(3)$
symmetry that acts on $n_{1},n_{2},n_{3}$. We have then constructed
a chiral spin liquid at the surface - the non-trivial quasiparticle
is a semion, and because it descends from a single vortex, it carries
spin $1/2$. The chiral spin liquid is of course a perfectly valid
2D symmetry enriched phase, which we expect since the bulk $SO(3)$
SPT is trivial.

Having understood this trivial case, we turn to $G=\mathbb{Z}_{2}^{A}\times\mathbb{Z}_{2}^{B}$
defined above, which corresponds to the non-trivial SPT. The analysis
is in fact completely the same, with the only difference being that
$\mathbb{Z}_{2}^{B}$ differs from $\tilde{\mathbb{Z}}_{2}^{B}$ by
the transformation $n_{4,5}\rightarrow-n_{4,5}$, i.e. a $\pi$ rotation
of the $U(1)$: $n_{4}+in_{5}\rightarrow-n_{4}-in_{5}$. Once we drive
the system into the $\nu=1/2$ bosonic Laughlin state, this extra
$\pi$ rotation causes $\mathbb{Z}_{2}^{B}$ to pick up an extra phase
factor, equal to $\pi$ times the charge of the semion (which is $1/2$),
i.e. $e^{i\pi/2}=i$. In other words, the semion now carries an extra
half charge of $\mathbb{Z}_{2}^{B}$, compared to the case of the
chiral spin liquid. This is precisely our anomalous surface state.
The other anomalous surface states can be constructed by changing
the roles of the generators of $\mathbb{Z}_{2}\times\mathbb{Z}_{2}$.

\section{Discussion and Future Directions}\label{disc}

In this paper, we studied in detail a set of symmetry enriched topological states -- the projective semion states -- which cannot be realized in 2D symmetric models even though the fractional symmetry action on the anyons is consistent with all the fusion and braiding rules of the chiral semion topological order. The anomaly is exposed when we try to gauge the $\mathbb{Z}_2 \times \mathbb{Z}_2$ symmetry and fail to find a solution for the fusion statistics of the gauge fluxes which satisfies the pentagon equation. On the other hand, we demonstrated that the projective semion state can be realized on the surface of a 3D system and proved that the anomaly determines the SPT order of the 3D bulk.

The projective semion state is the simplest example of an anomalous SET state with discrete unitary symmetries. Our discussion in section \ref{anomaly} illustrates a general procedure for detecting anomalies in a large class of such SET states. In particular, we want to emphasize two points:
\begin{enumerate}
\item The fractional symmetry action on the anyons gives rise to a projective fusion rule of the gauge fluxes once the symmetry is gauged, with the abelian anyon in the original topological theory as the coefficient. 
\item For SETs with discrete unitary symmetries, when gauging the symmetry, the possible obstruction comes in two types: the $H^3$ type and the $H^4$ type. Eq. \ref{ob} gives the general formula to detect the $H^4$ type. If the $\nu(f,g,h,k)$ in Eq. \ref{ob} forms a nontrivial 4-cocycle of $G$ with $U(1)$ coefficients, then the SET is anomalous.  Furthermore, $\nu(f,g,h,k)$ is a 4-cocycle representing the SPT order in the 3D bulk.
\end{enumerate}   

Thus far we have focused on discrete unitary symmetry groups $G$ and identifying whether a given SET is anomalous.  We also believe that there should be a constructive procedure similar to the one utilized here to realize a particular anomalous SET,  given an element of $H^4$, which is the surface topological order of the corresponding 3D bosonic SPT phase, at least for discrete unitary $G$. (A 3D SPT  with no consistent surface  topological order was recently discussed \cite{Wang2014}, although that involved bulk fermions and a combination of continuous symmetry and time reversal).  A construction similar to that of Sect. \ref{WW} for other abelian anyon models is discussed in Appendix \ref{gWW}.  

For time reversal symmetry, although it is currently not known how to introduce time reversal fluxes at the same level as for a unitary symmetry, we conjecture that the same formula Eq. \ref{ob} works in detecting the $H^4$ type obstruction. This comes from the observation of the following example: the $\mathbb{Z}_2$ gauge theory with both the gauge charge $e$ and gauge flux $m$ transforming as $T^2=-1$ (sometimes called eTmT). It is believed that this state is anomalous and appears on the surface of the 3D SPT with time reversal symmetry\cite{Vishwanath2013} which is described by the group cohomology classification, specifically by the nontrivial element in $H^4(Z_2^T,U(1))$, where the time reversal symmetry acts nontrivially on the $U(1)$ coefficients by complex conjugation. Suppose that we can define in some notion a time reversal gauge flux. Then the projective fusion rule would be
\be
\Om_T \times \Om_T = f
\ee
where $f$ is the bound particle of $e$ and $m$. Equivalently, we can write
\be
\omega(T,T)=f, \ \omega=I \ \text{otherwise}
\ee
This reflects the fractional symmetry action where $T^2=-1$ on $e$ and $m$ because $f$ has a $-1$ braiding statistics with both $e$ and $m$. Now we can use this information to calculate $\nu(f,g,h,k)$ in Eq. \ref{ob}. The $\mathbb{Z}_2$ gauge theory has trivial $F$. Therefore, Eq. \ref{ob} reduces to
\be
\nu(f,g,h,k) = R_{\omega(h,k),\omega(f,g)}
\ee
which is nontrivial only when $f=g=h=k=T$ and $\nu(T,T,T,T)=-1$. This is exactly the nontrivial 4-cocycle of time reversal \cite{Chen2013}. 

Even though at this moment, we are not sure what gauging time reversal means in general, recent work indicates that this notion can be formalized \cite{Chen2014,Kapustin2014}. From this particular example, we expect that the procedure and result discussed in Ref. \onlinecite{Etingof2010} might be generalized to treat  anti-unitary symmetries as well.
\section{acknowledgment}

When we were writing up the paper, we learned of other works on anomalous SET's with unitary discrete symmetries\cite{Kapustin2014,Cho2014}.  We also learned about the interesting work of reference \onlinecite{WangLevin} classifying 3D SPTs using flux loop braiding statistics; it would be nice to relate this to our results.

We are very grateful to helpful discussions with Meng Cheng, Senthil Todadri, Ryan Thorngren, Alexei Kitaev, Parsa Bonderson, and Netanel Lindner. XC is supported by the Miller Institute for Basic Research in Science at UC Berkeley, the Caltech Institute for Quantum Information and Matter and the Walter Burke Institute for Theoretical Physics. AV is supported by NSF DMR 1206728, and FB is supported by NSF DMR 1352271. 

\appendix

\section{Projective representation and group cohomology}
\label{Gcoh}

The following definition works only for unitary symmetries because we are not dealing with time reversal symmetry in this paper.

Matrices $u(g)$ form a projective representation of symmetry group $G$ if
\begin{align}
u(g_1)u(g_2)=\om(g_1,g_2)u(g_1g_2),\ \ \ \ \
g_1,g_2\in G.
\end{align}
Here $\om(g_1,g_2) \in U(1)$ and $\om(g_1,g_2) \neq 1$, which is called the
factor system of the projective representation. The factor system satisfies
\begin{align}
 \om(g_2,g_3)\om(g_1,g_2g_3)&=
 \om(g_1,g_2)\om(g_1g_2,g_3),
 \label{2cocycle_om}
\end{align}
for all $g_1,g_2,g_3\in G$.
If $\om(g_1,g_2)=1, \ \forall g_1,g_2$, this reduces to the usual linear representation of $G$.

A different choice of pre-factor for the representation matrices
$u'(g)= \bt(g) u(g)$ will lead to a different factor system
$\om'(g_1,g_2)$:
\begin{align}
\label{omom}
 \om'(g_1,g_2) =
\frac{\bt(g_1)\bt(g_2)}{\bt(g_1g_2)}
 \om(g_1,g_2).
\end{align}
We regard $u'(g)$ and $u(g)$ that differ only by a pre-factor as equivalent
projective representations and the corresponding factor systems $\om'(g_1,g_2)$
and $\om(g_1,g_2)$ as belonging to the same class $\om$.

Suppose that we have one projective representation $u_1(g)$ with factor system
$\om_1(g_1,g_2)$ of class $\om_1$ and another $u_2(g)$ with factor system
$\om_2(g_1,g_2)$ of class $\om_2$, obviously $u_1(g)\otimes u_2(g)$ is a
projective presentation with factor system $\om_1(g_1,g_2)\om_2(g_1,g_2)$. The
corresponding class $\om$ can be written as a sum $\om_1+\om_2$. Under such an
addition rule, the equivalence classes of factor systems form an Abelian group,
which is called the second cohomology group of $G$ and is denoted as
$\cH^2(G,U(1))$.  The identity element $1 \in \cH^2(G,U(1))$ is the class that
corresponds to the linear representation of the group.

The above discussion on the factor system of a projective representation can be
generalized which gives rise to a cohomology theory of groups.

For a group $G$, let $M$ be a G-module, which is an abelian group (with
multiplication operation) on which $G$ acts compatibly with the multiplication
operation (\ie the abelian group structure) on M:
\begin{align}
\label{gm}
 g\cdot (ab)=(g\cdot a)(g\cdot b),\ \ \ \ g\in G,\ \ \ \ a,b\in M.
\end{align}
For example, $M$ can be the $U(1)$ group and $a$ a
$U(1)$ phase. The multiplication operation $ab$ is then the usual multiplication of
the $U(1)$ phases. The group action is trivial $g\cdot a=a$ for unitary symmetries considered here. Or $M$ can be a $\mathbb{Z}_2$ group and $a$ is the semion or the vacuum sector in the $K=2$ Chern-Simons theory. The multiplication $ab$ is then the fusion between anyons. The group action $g\cdot a =b$ encodes how the anyon sectors get permuted under the symmetry, which is trivial for the projective semion example discussed in this paper but can be nontrivial in general.

Let $\om_n(g_1,...,g_n)$ be a function of $n$ group
elements whose value is in the G-module $M$. In other words, $\om_n:
G^n\to M$.  Let $\cC^n(G,M)=\{\om_n \}$ be the space of all such
functions.
Note that $\cC^n(G,M)$ is an Abelian group
under the function multiplication
$ \om''_n(g_1,...,g_n)= \om_n(g_1,...,g_n) \om'_n(g_1,...,g_n) $.
We define a map $d_n$ from $\cC^n(G,U(1))$ to $\cC^{n+1}(G,U(1))$:
\begin{align}
&\ \ \ \
(d_n \om_n) (g_1,...,g_{n+1})=
\nonumber\\
&
g_1\cdot \om_n (g_2,...,g_{n+1})
\om_n^{(-1)^{n+1}} (g_1,...,g_{n}) \times
\nonumber\\
&\ \ \ \ \
\prod_{i=1}^n
\om_n^{(-1)^i} (g_1,...,g_{i-1},g_ig_{i+1},g_{i+2},...g_{n+1})
\end{align}
Let
\begin{align}
 \cB^n(G,M)=\{ \om_n| \om_n=d_{n-1} \om_{n-1}|  \om_{n-1} \in \cC^{n-1}(G,M) \}
\end{align}
and
\begin{align}
 \cZ^n(G,M)=\{ \om_{n}|d_n \om_n=1,  \om_{n} \in \cC^{n}(G,M) \}
\end{align}
$\cB^n(G,M)$ and $\cZ^n(G,M)$ are also Abelian groups
which satisfy $\cB^n(G,M) \subset \cZ^n(G,M)$ where
$\cB^1(G,M)\equiv \{ 1\}$. $\cZ^n(G,M)$ is the group of $n$-cocycles and
$\cB^n(G,M)$ is the group of $n$-coboundaries.
The $n$th cohomology group of $G$ is defined as
\begin{align}
 \cH^n(G,M)= \cZ^n(G,M) /\cB^n(G,M)
\end{align}

When $n=1$, we find that $\om_1(g)$ satisfies
\be
\om_1(g_1)\om_1(g_2)=\om_1(g_1g_2)
\ee
Therefore, the 1st cocycles of a group with $U(1)$ coefficient are the one dimensional representations of the group.

Moreover, we can check that the consistency and equivalence conditions (Eq. \ref{2cocycle_om} and \ref{omom}) of factor systems of projective representations are exactly the cocycle and coboundary conditions of 2nd cohomology group. Therefore, 2nd cocycles of a group with $U(1)$ coefficient are the factor systems of the projective representations of the group. Similarly, we can check that if we use the semion / vacuum sector as a $Z_2$ coefficient, then the projective fusion rule of the gauge fluxes discussed in section \ref{proj_fusion} is a 2nd cocycle of the symmetry group with $Z_2$ coefficient. 

When $n=3$, from
\begin{align}
&\ \ \ \ (d_3 \om_3)(g_1,g_2,g_3,g_4)
\nonumber\\
&= \frac{ \om_3(g_2,g_3,g_4) \om_3(g_1,g_2g_3,g_4)\om_3(g_1,g_2,g_3) }
{\om_3(g_1g_2,g_3,g_4)\om_3(g_1,g_2,g_3g_4)}
\end{align}
we see that
\begin{align}
& \cZ^3(G,U(1))=\{  \om_3|
\\
&\ \ \ \frac{ \om_3(g_2,g_3,g_4) \om_3(g_1,g_2g_3,g_4)\om_3(g_1,g_2,g_3) }
{\om_3(g_1g_2,g_3,g_4)\om_3(g_1,g_2,g_3g_4)}
=1
 \} .
\nonumber
\end{align}
and
\begin{align}
& \cB^3(G,U(1))=\{ \om_3| \om_3(g_1,g_2,g_3)=\frac{
\om_2(g_2,g_3) \om_2(g_1,g_2g_3)}{\om_2(g_1g_2,g_3)\om_2(g_1,g_2)}
 \},
\label{3coboundary}
\end{align}
which give us the third cohomology group
$\cH^3(G,U(1))=\cZ^3(G,U(1))/\cB^3(G,U(1))$.

Similarly, when $n=4$, from
\begin{align}
&\ \ \ \ (d_4 \om_4)(g_1,g_2,g_3,g_4,g_5)
\nonumber\\
&= \frac{ \om_4(g_2,g_3,g_4,g_5) \om_4(g_1,g_2g_3,g_4,g_5)\om_4(g_1,g_2,g_3,g_4g_5) }
{\om_4(g_1g_2,g_3,g_4,g_5)\om_4(g_1,g_2,g_3g_4,g_5)\om_4(g_1,g_2,g_3,g_4) }
\end{align}
we see that
\begin{align}
& \cZ^4(G,U(1))=\{  \om_4|
\\
&\frac{ \om_4(g_2,g_3,g_4,g_5) \om_4(g_1,g_2g_3,g_4,g_5)\om_4(g_1,g_2,g_3,g_4g_5) }
{\om_4(g_1g_2,g_3,g_4,g_5)\om_4(g_1,g_2,g_3g_4,g_5)\om_4(g_1,g_2,g_3,g_4) }
=1
 \} .
\nonumber
\end{align}
and
\begin{align}
& \cB^4(G,U(1))=\{ \om_4| \om_4(g_1,g_2,g_3,g_4)= \\
& \frac{
\om_3(g_2,g_3,g_4) \om_3(g_1,g_2g_3,g_4)\om_3(g_1,g_2,g_3) }
{\om_3(g_1g_2,g_3,g_4)\om_3(g_1,g_2,g_3g_4)}
 \},
\label{4coboundary}
\end{align}
which give us the third cohomology group
$\cH^4(G,U(1))=\cZ^4(G,U(1))/\cB^4(G,U(1))$.

\section{Showing that the 4-cocycle we get is non-trivial}
\label{4cocycle}

Although we have numerically verified that the 4-cocycle we obtained
from the pentagon anomaly argument corresponds to a non-trivial cohomology
class in $H^{4}(G,U(1))$, it is nice to also have an analytic proof
of this fact. This is what we demonstrate in this appendix. The notation
is: $[\nu]$ is the $H^{4}(G,U(1))$ obstruction class, and $[\omega_{1}],[\omega_{2}]$
are $H^{2}(\mathbb{Z}_{2}\times\mathbb{Z}_{2},\mathbb{Z}_{2})$ classes.
The former is computed from the latter, and in fact Ref.\onlinecite{Etingof2010}
show that:

\begin{center}
$[\nu]([\omega_{1}]+[\omega_{2}])=[\nu]([\omega_{1}])\,[\nu]([\omega_{2}])\, b([\omega_{1}],[\omega_{2}])$
\par\end{center}

\noindent where the ``bilinear form'' $b$ is defined by taking
the full braid of $\omega_{1}(f,g)$ with $\omega_{2}(g,h)$. For
our purposes $[\omega_{1}]$ will correspond to the chiral spin liquid
(which is not anomalous), and $[\omega_{2}]$ will correspond to a
theory where the semion only binds a half-charge of some $\mathbb{Z}_{2}\subset\mathbb{Z}_{2}\times\mathbb{Z}_{2}$.
It is useful to let $u_{1}$ and $u_{2}$ denote elements of $H^{1}(\mathbb{Z}_{2},\mathbb{Z}_{2})$
which correspond to the first and second $\mathbb{Z}_{2}$. Then (using
multiplication in the cohomology ring) we can have $[\omega_{2}]=u_{1}^{2},u_{2}^{2},$
or $u_{1}^{2}+u_{2}^{2}$, whereas $[\omega_{1}]$ will be denoted
$v$, the class of the chiral spin liquid. We know that both the CSL
and the theory corresponding to $[\omega_{2}]$ are realizable in
2d, so their obstructions vanish: $[\nu](v)=[\nu](u_{1}^{2})=[\nu](u_{2}^{2})=1$.
So we see that the obstruction correspoding to $v+\alpha_{1}u_{1}^{2}+\alpha_{2}u_{2}^{2}$
is simply $(b(v,u_{1}^{2}))^{\alpha_{1}}(b(v,u_{2}^{2}))^{\alpha_{2}}$,
where $\alpha_{i}=0,1$. Now, in the case of our semion theory, the
full braid of two anyons gives $-1$ only when both anyons are equal
to $s$; if we view the anyons $\mathbb{Z}_{2}=\{0,1\}$ where $0$
is the trivial anyon and $1$ is $s$, then this is just multiplication
in the field $\mathbb{Z}_{2}$. Hence $b(v,u_{i}^{2})$ is simply
the product of $u_{i}^{2}$ and $v$ in the cohomology ring with $\mathbb{Z}_{2}$
coefficients. This is $u_{1}^{3}u_{2}$ in the case $i=1$ and $u_{1}u_{2}^{3}$
in the case $i=2$. In particular, both of these are nonzero when
viewed as elements of $H^{4}(G,\mathbb{Z}_{2})$: this is because,
according to the Kunneth formula, $H^{4}(G,\mathbb{Z}_{2})$ is spanned
as a $\mathbb{Z}_{2}$ vector space by $\{u_{1}^{4},u_{1}^{3}u_{2},u_{1}^{2}u_{2}^{2},u_{1}u_{2}^{3},u_{2}^{4}\}$.

The only thing we now need to check is that both $u_{1}^{3}u_{2}$
and $u_{1}u_{2}^{3}$ are nonzero as $H^{4}(G,U(1))$ cohomology classes.
In fact, $H^{4}(G,U(1))=\mathbb{Z}_{2}\times\mathbb{Z}_{2}$, and
the kernel of the natural map $H^{4}(G,\mathbb{Z}_{2})\rightarrow H^{4}(G,U(1))$
is precisely $\{u_{1}^{4},u_{1}^{2}u_{2}^{2},u_{2}^{4}\}$. To prove
this, first note that the map $H^{4}(G,\mathbb{Z}_{2})\rightarrow H^{4}(G,U(1))$
is equivalent to the coboundary map $H^{4}(G,\mathbb{Z}_{2})\rightarrow H^{5}(G,\mathbb{Z})$
from the long exact sequence associated to $\mathbb{Z}\rightarrow\mathbb{Z}\rightarrow\mathbb{Z}_{2}$,
under the natural identification of $H^{4}(G,U(1))$ and $H^{5}(G,\mathbb{Z})$.
Then we use the explicit Kunneth formula for cohomology with $\mathbb{Z}$
coefficients, which contains both the ``normal'' part that is already
visible with $\mathbb{Z}_{2}$ coefficients, and the torsion part
in one degree higher, and the fact that the torsion part maps into
the normal part under the coboundary map, to conclude that the kernel
of $H^{4}(G,\mathbb{Z}_{2})\rightarrow H^{4}(G,U(1))$ is precisely
$\{u_{1}^{4},u_{1}^{2}u_{2}^{2},u_{2}^{4}\}$.

\section{Generalization of the Walker-Wang model to other cases}
\label{gWW}
  
 The Walker-Wang semion model discussed in section \ref{WW} is only one member of a family of models that have no deconfined excitations in the bulk, but are topologically ordered on the surface\cite{Walker12,vonkeyserlingk13}.  When this topological order is abelian, we can use a construction analogous to the one presented in Sect. \ref{WW} to  decorate the model with global symmetries such that the surface anyons transform projectively.  Such a construction can realize a surface state in which anyons transform under any of the projective representations consistent with their fusion rules, and hence can realize both anomalous and non-anomalous surface states.  Here we will sketch the main features of this construction.  
 
 A more general Walker-Wang model consists of an $n$-state system on each edge of the (point-split) cubic lattice.  In the models of interest to us, each of these $n$ states can be associated with an anyon type that will appear at the surface.   In general, to fix the state on a particular edge we must also fix an orientation of this edge; the label $a$ can then be viewed as a flux along the edge in question, whose direction is specified by this orientation.  The flux in the opposite direction is given by the conjugate anyon label $\overline{a}$. 
 
The Walker-Wang Hamiltonian has the general form (\ref{WWHam}), with commuting vertex and plaquette operators.  The plaquette term is discussed in detail in Refs. \onlinecite{Walker12,vonkeyserlingk13}; for our purposes it is sufficient to know that (provided our anyon model is modular) this plaquette operator confines all open strings in the bulk, but allows deconfined anyons at the surface.
To specify the vertex operator, we must use the fusion rules of the anyon model, which dictate how anyons proximate in space can combine to give new anyons (or possibly the trivial state $\ket{0}$, with no anyons at all).   
 The vertices $V$ that are allowed in the ground state are those where the anyons entering $V$ can be fused to each other to give $\ket{0}$.  For example, a vertex with $a$ entering on one edge, $\overline{a}$ entering (or equivalently, $a$ exiting)  on another edge, and all other edges in the state $\ket{0}$ is always allowed.  (For the 2-state model described above, this is the only allowed vertex).  However, in general there will be other allowed vertex types as well.  
  
 The simplest example of this is a $\mathbb{Z}_3$ model in which $\overline{0} = 0, \overline{1} = 2, \overline{2} =1$, and $1$ fuses with $2$ to give $0$.  The vertices allowed in the ground state (on the point-split lattice, where all vertices are trivalent) are ones in which a $1$ (or $2$) anyon flux enters the vertex on one edge, and exits on another, or vertices at which 3 anyons fluxes of the same type all enter.

How do we decorate this model such that each anyon transforms in a particular representation of our global symmetry group at the surface?  We first enlarge the Hilbert space at each (trivalent) vertex by including 3 auxillary sites (one associated to each edge entering the vertex).  The Hilbert space of each auxillary site contains a set of objects transforming in different representations of $G$.  In our example above, this set contained an object $b$ transforming in the singlet representation, and a spin $1/2$ (transforming, unsurprisingly, in the spin-$1/2$ representation).   More generally, let $r_a$ be the representation that we wish to make the anyon $a$ transform in.  Then our auxillary Hilbert space must contain an object transforming in $r_a$ for every $a$.  For notational simplicity, we will call this object $r_a$ from now on.  

In order for our construction to work, we must impose the condition that our choice of $r_a$ is consistent with the fusion rules of the anyon model.  Specifically, we will require that $r_0$ is the trivial representation, and that  conjugate anyon types transform under conjugate representations in the symmetry group:
\be
\overline{r_a} = r_{\overline {a}} 
\ee
We also require that 
\be \label{LinConstr}
a \times b = c   \ \ \ \Rightarrow \ \ \ r_a \times r_b = r_c  \times \text{  linear reps}
\ee
(For non-abelian anyons, $c$ may contain more than one anyon type.  However, we will require the projective part of the symmetry action to be the same for all possible products of fusion.)  

The next step is to construct analogues of the states $\ket{\tilde{0}}$ and $\ket{\tilde{1}}$ for this more general case.
Because the edges are oriented, it is natural to favour configurations in which the edge starting at vertex $V_1$ and ending at vertex $V_2$, with anyon label $a$, has auxillary variables in $r_a$ at $V_1$ and in $\overline{r_a}$ at $V_2$.  The analogue of the two states $\ket{1}$ and $\ket{0}$ above are states in which an edge with  anyon label $a$ has these two auxillary variables combining to the trivial representation (i.e., in which $r_a$ and $\overline{r}_a$ are in a ``singlet" state):
\be
\ket{\tilde{a}} = \ket{a} \ket{ P_{r_a \times \overline{r_a} = r_0 } } \ \ \ .
\ee
 We may favour such configurations energetically by introducing a potential term
 \be
 H_0 = \sum_{e = \text{edges} } \sum_{a=1}^{n} \ket{ \tilde{a}_e } \langle \tilde{a}_e |
 \ee

We next impose the constraint that only linear representations are allowed at vertices.  In the ground state, the tensor product of the representations associated with the anyon types that meet at each vertex contains the identity, so Eq. (\ref{LinConstr}) ensures that this constraint is compatible with minimizing $H_0$ on each edge.  At vertices where the net anyon flux is not zero, however, this constraint forces us to include at least one edge on which $H_0$ is violated.  Let the anyon types of the 3 edges (all oriented into the vertex) be $a,b$ and $c$, and let them be fused at the vertex to create a fourth anyon $d$.  Let us choose the edge with label $a$ to be an excited state of $H_0$.  Then edges $b$ and $c$ contribute $\overline{r}_b \times \overline{r}_c$ to the vertex; hence (up to a tensor product with linear representations, which is not important for our purposes) the auxillary variable associated with $a$ must carry a representation of $r_b \times r_c$.  It follows that the edge $a$ has two auxillary variables, which together transform in the representation $r_a \times r_b \times r_c$.  (Recall that the other end of our $a$-labelled edge had better carry $r_a$, to avoid having edges at adjacent vertices that are also in excited states).  But we have required that (up to a tensor product with linear representations), $r_a \times r_b \times r_c = r_d$.  Hence a vertex with net anyon flux $d$ necessarily has the correct projective component of its symmetry transformation.  (In cases where this leaves the representation under which $d$ transforms ambiguous, it is possible to add additional potential terms to remove this ambiguity).

Evidently, with this construction it is possible to define operators that mix the $\ket{\tilde{a}}$ variables in a manner consistent with the symmetry, since for all states $\ket{\tilde{a}}$ the edge carries a trivial representation of the symmetry.  We may thus as above construct the plaquette term of the Walker-Wang Hamiltonian in the $\tilde{a}$ variables to obtain a model in which anyons are confined in the bulk, and transform in the desired projective representations on the surface.

\end{document}